\documentclass[aip,apl,reprint]{revtex4-2}
\usepackage{dcolumn}
\usepackage{bm}
\usepackage[utf8]{inputenc}
\usepackage[T1]{fontenc}
\usepackage{mathptmx}
\usepackage{xcolor}
\usepackage{amsmath}
\usepackage{gensymb}
\usepackage{textcomp}
\usepackage{epstopdf}
\usepackage{graphicx,dcolumn,bm,color,bbding,epstopdf}

\draft 

\begin{document}
	
	
	
	
	
	%
	%
	
	%
	
	
	\title[Coherent length in SHG microscopy]{Quantifying the Coherent Interaction Length of Second-Harmonic Microscopy in Lithium Niobate Confined Nanostructures} 
	
	
	
	\author{Zeeshan H. Amber}
	\affiliation{TU Dresden, Institute of Applied Physics, Nöthnitzer Strasse 61, 01187 Dresden, Germany}
	
	\author{Benjamin Kirbus}
	\affiliation{TU Dresden, Institute of Applied Physics, Nöthnitzer Strasse 61, 01187 Dresden, Germany}
	
	\author{Lukas M. Eng}
	\affiliation{TU Dresden, Institute of Applied Physics, Nöthnitzer Strasse 61, 01187 Dresden, Germany}
	\affiliation{ct.qmat: Dresden-Würzburg Cluster of Excellence—EXC 2147, TU Dresden, 01062 Dresden, Germany}
	
		\author{Michael Rüsing}
		 \email{michael.ruesing@tu-dresden.de}
	\affiliation{TU Dresden, Institute of Applied Physics, Nöthnitzer Strasse 61, 01187 Dresden, Germany}

	
	\date{\today}
	
	\begin{abstract}
		Thin-film lithium niobate (TFLN) in the form of x- or z-cut lithium-niobate-on-insulator (LNOI) has attracted considerable interest as a very promising and novel platform for developing integrated optoelectronic (nano)devices and exploring fundamental research. Here, we investigate the coherent interaction length $l_{c}$ of optical second-harmonic generation (SHG) microscopy in such samples, that are purposely prepared into a wedge shape, in order to elegantly tune the geometrical confinement from bulk thicknesses down to approximately 50 nm. SHG microscopy is a very powerful and non-invasive tool for the investigation of structural properties in the biological and solid-state sciences, especially also for visualizing and analyzing ferroelectric domains and domain walls. However, unlike in bulk LN, SHG microscopy in TFLN is impacted by interfacial reflections and resonant enhancement which both rely on film thickness and substrate material. In this paper we show that the dominant SHG contribution measured on TFLN in back-reflection, is the co-propagating phase-matched SHG signal and \textit{not} the counter-propagating SHG portion as is the case for bulk LN samples. Moreover,  $l_{c}$ depends on the incident pump laser wavelength (sample dispersion) but also on the numerical aperture of the focussing objective in use. These experimental findings on x- and z-cut TFLN are excellently backed up by our advanced numerical simulations. 
	\end{abstract}

	
	\maketitle 
	
	
	\section{Introduction}
	\label{sec_intro}
	
	Nonlinear optical (NLO) microscopy techniques such as second-harmonic generation (SHG), third-harmonic generation (THG), or coherent Anti-stokes Raman spectroscopy (CARS), see widespread use for both imaging and the spectroscopic analysis from cell biology to material sciences \cite{Squier2001,Cox2011,Kumar2013,Carriles2009,Oron2004}. All these methods rely on the coherent NLO interaction of electromagnetic waves at a single or multiple fundamental frequencies, in order to generate a NLO signal at a different frequency. The NLO-generation efficiencies strongly depend on the underlying, local material properties. Conversely, they equally provide a label-free contrast mechanism for an in-depth analysis of distinct material properties using NLO imaging. For example, CARS spectroscopy is sensitive to molecular vibrations and phonons in solids, thus fingerprinting the local chemical composition or local structural changes \cite{Golovan2012,Day2011,Djaker2007}. Bulk SHG, in contrast, works only in non-centrosymmetric materials (e.g. piezoelectrics and ferroelectrics) and strongly depends on the material symmetry and incident light polarization \cite{Cherifi-Hertel2017,Cherifi-Hertel2021,Rusing2019e,Spychala2017,Reitzig2021}. SHG polarimetry thus witnesses a widespread use to analyze the orientation and crystallographic structure of appropriate (nano)systems. In this context, SHG might be even applied to study phase transitions, to visualize ferroelectric (FE) domains and domain walls (DWs) therein \cite{Berth2007,Kampfe2014,Kampfe2015,Kurimura1997,Florsheimer1998}, or to analyze novel topologies\cite{Cherifi-Hertel2017,Cherifi-Hertel2021}.
	
	NLO effects \textit{per se} are coherent optical processes. In the absence of dispersion, we witness a monotonic and nonlinear increase of the generated NLO signal as a function of interaction length. However, due to dispersion, each of the interacting waves propagates with different velocities, inherently resulting in an increasing phase difference between the fundamental and NLO signal beams. This limits the overall efficiency of any NLO process and is commonly referred to as the \textit{phase-matching} problem. In this context, the coherent interaction length $l_{c}$ specifies exactly the length up to which a NLO process is still constructive \cite{Boyd2003}. Notably, the overall NLO signal amplitude will oscillate with twice the length $l_{c}$.
	
	\begin{figure}
		\centering
		\includegraphics[width=0.95\linewidth]{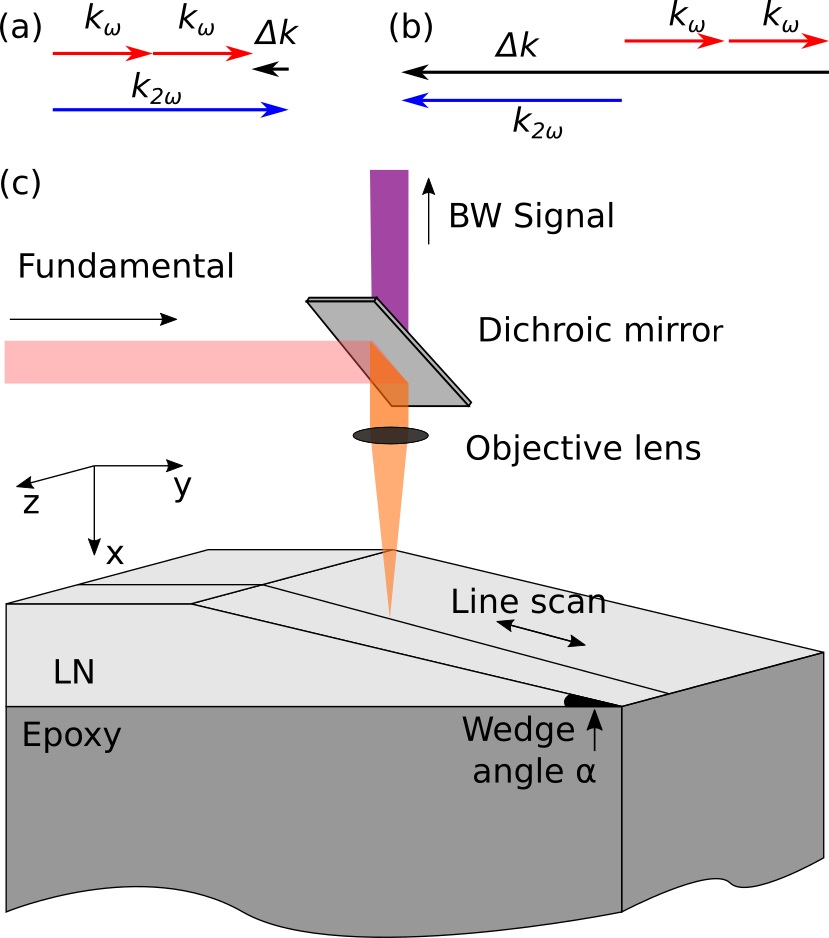}
		\caption{Vector diagrams for the SH generation process for (a) co- and (b) counter-propagating beams; (c) Schematic drawing of the experimental geometry. Here, the incident beam $\omega_{inc} = \omega$ is scanned across the LiNbO$_{3}$ (LN) wedge while recording the $\omega_{SHG} = 2\omega$ signal in back reflection.}
		\label{fig 1}
	\end{figure} 
	
	SHG is the archetypical NLO process where only two waves are involved, the fundamental (pump) beam at frequency $\omega_{inc} = \omega$ and the SHG wave $\omega_{SHG} = 2\omega$. Phase matching of these two waves in $k$-space requires 
	\begin{equation}
		\label{eqn:1}
		2k_{\omega} - \Delta k = \pm k_{2\omega} .
	\end{equation}
	with $\Delta k$ the  phase mismatch. Here, two main processes may be differentiated (see Fig. 1), (i) co-propagation with $+k_{2\omega}$ being collinear, and (ii) counter-propagation with $-k_{2\omega}$ aligning anti-collinearly with respect to the fundamental wave vector $k$, respectively. Using $k=2{\pi}n_{\omega}/{\lambda_{\omega}}$, with $n_{\omega}$ the refractive index at the given $\lambda_{\omega}$, and ${\Delta}k = {\pi}/l_c$, we obtain for $l_c$: 
	\begin{equation}
		\label{eqn:2}
		l_c = \frac{\lambda_{\omega}}{4}\frac{1}{n_{2\omega} \pm n_{\omega}}.
	\end{equation}
	Note the positive $/$ negative sign in $k_{2\omega}$ in Eq. \ref{eqn:2}, standing for the counter- and co-propagating SH wave, respectively. To underline the importance of such a differentiation between (i) and (ii), a quick calculus for the lithium niobate (LN) sample used here, pumped at 800 nm (using a Ti:Sapphire laser) reveals the following two interaction lengths for co- and counter propagating waves: $l_{c,co} \approx 1280$~nm and $l_{c, counter} \approx 44$~nm. Since SHG intensities scale quadratically with interaction length, co-propagating SH light intensities are approximately three orders of magnitude larger as compared to the counter-propagating SH light intensities (assuming a sample size of exactly $l_c$). Similar arguments can be made for other NLO processes, such as THG or CARS.
	
	Experimentally, co- and counter-propagating light in general is easily separated in SHG microscopy, due to their opposite $k$-vector behavior. Nevertheless, when mounting the sample under inspection on a substrate support (as is usually necessary when investigating thinner films or 2D materials)\cite{Kumar2013,Rusing2019e}, a significant amount of the co-propagating light can be reflected off the sample/substrate interface, and thus scatters into backward direction. Due to the potentially huge intensity difference between the co- and counter-propagating waves even a few percent of reflectivity can amount to a significant additional signal strength in backward-direction. Even more importantly, reflective interfaces may also resonantly enhance both the fundamental and/or the SH beams, hence additionally complicating experimental interpretation.
	
	Note that the above concerns might have negligible impact when applying standard NLO microscopy for instance to cell biology investigations. Firstly, active layers might be much smaller in size than the coherent interaction length $l_c$, while secondly, only few and weakly reflective interfaces are present that might cause back-reflection or interface enhancement. In modern-type material science, however, a SHG-microscopy investigation of nanostructures and samples with confining structures is of utmost interest, hence automatically needing to address the above issues. Therefore, effects of reflection, resonant enhancement, and/or different phase matching scenarios, all need to be carefully taken into account when aiming at the quantitative analysis of the measured SHG contrast. 
	
	In the work presented here, we systematically investigate these effects by comparing our experimental findings from SHG microscopy to our rigorous numerical simulations. Both x- and z-cut LN wedges mounted on a Si and epoxy substrate, respectively, are analyzed, allowing us to elegantly change the sample thickness $t$ from tens of nanometers up to the bulk threshold, through simply scanning the incident linearly polarized photon beam across the wedges' slope (see Fig. 1). Our study paves the way for the systematic understanding of both reflections and the role of phase matching when applying NLO microscopy to (ultra)thin films and nanostructures, i.e. ferroelectric and ferroic thin films, or even 2D materials. Due to the archetypical nature of the SHG process, these findings can readily be translated to any other NLO process.

	\section{METHODOLOGY AND PHYSICAL PRINCIPLE}
	\label{sec_METHODOLOGY AND PHYSICAL PRINCIPLES}
	
		We use single-crystalline 5-$\%$MgO-doped congruent LN for our SHG studies here, prepared into wedges as x- and z-cut thin films and mounted onto an epoxy and Si substrate, respectively (for details see Supplementary Information I.B.). LN is one of the most common NLO material and sees widespread use in electronics as surface-acoustic wave filters, high-frequency electro-optical modulators, or in quantum optics. Moreover, the current boost in implementing LN into CMOS-compatible architectures can undoubtably be attributed to thin-film lithium niobate (TFLN) that provides many benefits as compared to the standard bulk LN platform \cite{Rusing2019b,Hu2009,Weigel2018d,Bartasyte2017,Rao2018,Sohler2008,Zhao2020,Wang2018f}. LN in the form as prepared here into wedges, hence, provides an ideal test system for quantifying the coherent interaction length $l_{c}$ and therefore analyzing the phase-matching process as needed in many NLO applications. 
			
		SHG scattering is described through the induced SHG polarization $\vec{P}$, with $I_{SH} \propto \vec{P}^2$ the SH intensity. $\vec{P}$ is readily calculated through 
	\begin{equation}
		\vec{P} = \begin{pmatrix} P_{x}\\P_{y}\\P_{z} \end{pmatrix}
		=
		{\chi}^{(2)}
		\begin{pmatrix}
			{E_{x}}^2\\{E_{y}}^2\\{E_{z}}^2\\2E_{z}E_{y}\\2E_{z}E_{x}\\2E_{x}E_{y} 
		\end{pmatrix},
	\end{equation}
		with $E_x$, $E_y$, and $E_z$ denoting the electric field components at the fundamental wavelength along the crystal axes x,y,z, respectively, and $\chi^{(2)}$ being the second order NLO susceptibility tensor given for LN as: 	
	\begin{equation}
		{\chi}^{(2)} = 
		\begin{pmatrix}
			0	&	0	&	0	&	0	&	d_{15}	&	-d_{16}\\
			-d_{21}	&	d_{22}	&	0	&	d_{24}	&	0	&	0\\
			d_{31}	&	d_{32}	&	d_{33}	&	0	&	0	&	0
		\end{pmatrix}.
	\end{equation}
	
\begin{figure}
	\centering
	\includegraphics[width=0.95\linewidth]{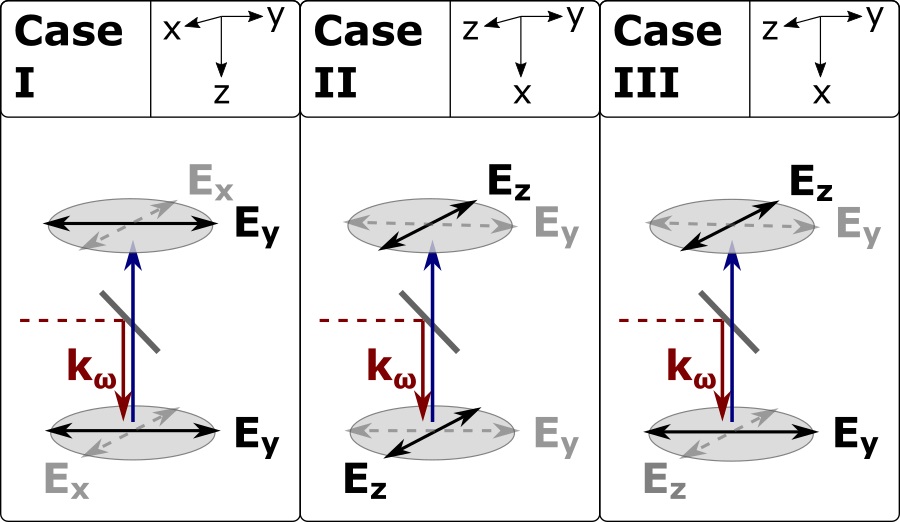}
	\caption{Schematic drawing of the three experimental scenarios.}
	\label{Fig:cases}
\end{figure}

		Due to symmetry, only three components $d_{ij}$ are independent in LN, i.e.  $d_{21} = d_{22} = d_{16} = 3$~pm/V, $d_{31} = d_{32} = d_{24} = d_{15} = 6$~pm/V, and $d_{33} = 36$~pm/V, which can be assumed approximately constant over the relevant wavelength range from 850 to 950 nm \cite{Nikogosjan2005,Shoji1997,Riefer2013b}. Moreover, LN is an uniaxial birefringent material, and hence features two independent refractive indices, the ordinary and extraordinary refractive indices $n_o$ and $n_e$ for light polarized in the x-y plane and along the crystallographic z-axes, respectively \cite{Zelmon2008}. Combining these two refractive indices with the SHG susceptibility tensor  $d_{ij}$, we identify three distinct scenarios as listed in Tab. \ref{cases} and schematically depicted in Fig. \ref{Fig:cases}, in order to investigate $l_c$ by SHG microscopy along the x- and z-cut LN wedges:
		
	\begin{itemize}
		\item[i)] Case I: When using z-cut TFLN, the linearly-polarized incident beam propagates along the z-axis [see Fig. \ref{Fig:cases}(a)], with hence the backreflected SHG light stemming from the SHG coefficients $d_{21}$ or $d_{22}$ within the x-y-plane. Since both parameters yield similar results (see Supplementary Information Section II.), we consider here the case of y-polarized light, only. Due to strong dispersion, $l_c$ will be relatively small. 
		\item[ii)] Case II: For x-cut TFLN with the incident x-aligned beam being polarized along the extraordinary z-axes [see Fig. \ref{Fig:cases}(b)], the SHG response is exclusively generated through the largest tensor element $d_{33}$ and polarized along the extraordinary axes. We thus expect a slightly larger coherent interaction length as compared to case I. 
		\item[iii)] Case III: Here, the incident beam hits the x-cut TFLN under ordinary polarization (y-direction) while SHG light leaves the crystal z-polarized (extraordinary polarization), as mediated via the $d_{32}$ element [see Fig. \ref{Fig:cases}(c)]. Due to LN's strong birefringence, an almost 3$\times$ larger $l_c$ results, that converges to infinity for pump wavelengths close to 1080~nm for 5\%MgO:LN \cite{Kaneshiro2008}. This is the  well-known case of birefringence-phase matching in NLO, and hence widely exploited experimentally when fabricating highly-efficient NLO devices.
	\end{itemize}

\begin{table}
	\begin{tabular}{c|c|c|c|c|c}	
	 Case	& Crystal	& Phase matching& SHG Signal 	 & $l_{c,co}$  & substrate\\
			& cut	& process 			&dominated by 		&  @850 nm & \\
			\hline

	I 		&	 z-cut & $n_o(\omega) \rightarrow n_o(2\omega)$ &$P_y = d_{22}E^2_y$ &  1.36 $\mu$m & Si\\
		
	II &	x-cut &  $n_e(\omega) \rightarrow n_e(2\omega)$ & $P_z = d_{33}E^2_z$ & 1.60 $\mu$m & epoxy\\
		
	III &	x-cut & $n_o(\omega) \rightarrow n_e(2\omega)$ & $P_y = d_{32}E^2_y$ &  3.95 $\mu$m & epoxy\\
		
	\end{tabular}
	\caption{Investigated experimental scenarios and relevant details on the SHG processes.}	
	\label{cases}
\end{table}
			
	To study the coherence length for the three cases listed in Tab. \ref{cases}, wedge-samples have been fabricated in collaboration with SGS Fresenius GmbH as shown in Fig. \ref{fig 1} from z- and x-cut LN, respectively. The z-cut sample was bonded to a (001)-silicon wafer, providing an enhanced stability that allowed wedge polishing to a smaller angle. For polishing, both samples were completely immersed in epoxy and polished until interference patterns became visible at the thinner end of the wedges. indicating sub-wavelength thicknesses. The thickness and angle of the wedges were determined via Confocal laser scanning microscopy at a 405~nm wavelength (Olympus LEXT OLS4000). The inclination angle $\alpha$ of the z- and x-cut sample was determined over the first 2~mm to $\alpha_z = 0.45$\textdegree\ $ \pm 0.01$\textdegree\ and $\alpha_x = 1.97$\textdegree\ $\pm 0.01$\textdegree, respectively, while the wedge surface roughness evaluated from optical measurement was less than $<25$~nm$_{RMS}$. Details on the fabrication process and angle extraction can be found in the Supplemental Information sections I.B and I.C.
	
	All SHG experiments were performed on a commercial microscope (Zeiss LSM980) equipped with a tunable Ti:Sapphire laser light source (see Supplemental Information section  I.A). In our experiments, we employed two different objective lenses with numerical apertures NA~=~0.45 and NA~=~0.8. The incident pump light is always linearly polarized while the back-reflected SHG light contains all polarizations (unpolarized). Note, that each of the above discussed three cases is dominated by one major SHG process as denoted in Tab. \ref{cases}, making polarization analysis obsolete. Moreover, this is also confirmed with our numerical calculations. 
		
	Numerical calculations based on a code previously developed by Sandkujil et al. were carried out to complement our experimental findings \cite{Sandkuijl2013,Sandkuijl2013a,Rusing2019e,Spychala2020a,Sandkuijl}. These simulations feature a full 3-dimensional vectorial model for both, the focusing and collection processes, as well as a matrix formalism in order to describe influences of (multiple) reflections and/or (self) interferences occurring for focal beams. The model hence even accounts for larger k-vector spreads when using high-NA objectives; for example, this allows to address and quantify the impact of mixed polarization components $d_{24}$ or $d_{15}$ on the SHG efficiency by solely using linearly-polarized incident light  \cite{Spychala2020a,Spychala2020}. We thus performed all calculations with a linear polarization input and using the full $\chi^{(2)}$-tensor. Details on our numerical simulations can be found in the Supplement (section I.D).

	\section{Results}
	
		\begin{figure*}
		\centering
		\includegraphics[width=0.95\textwidth]{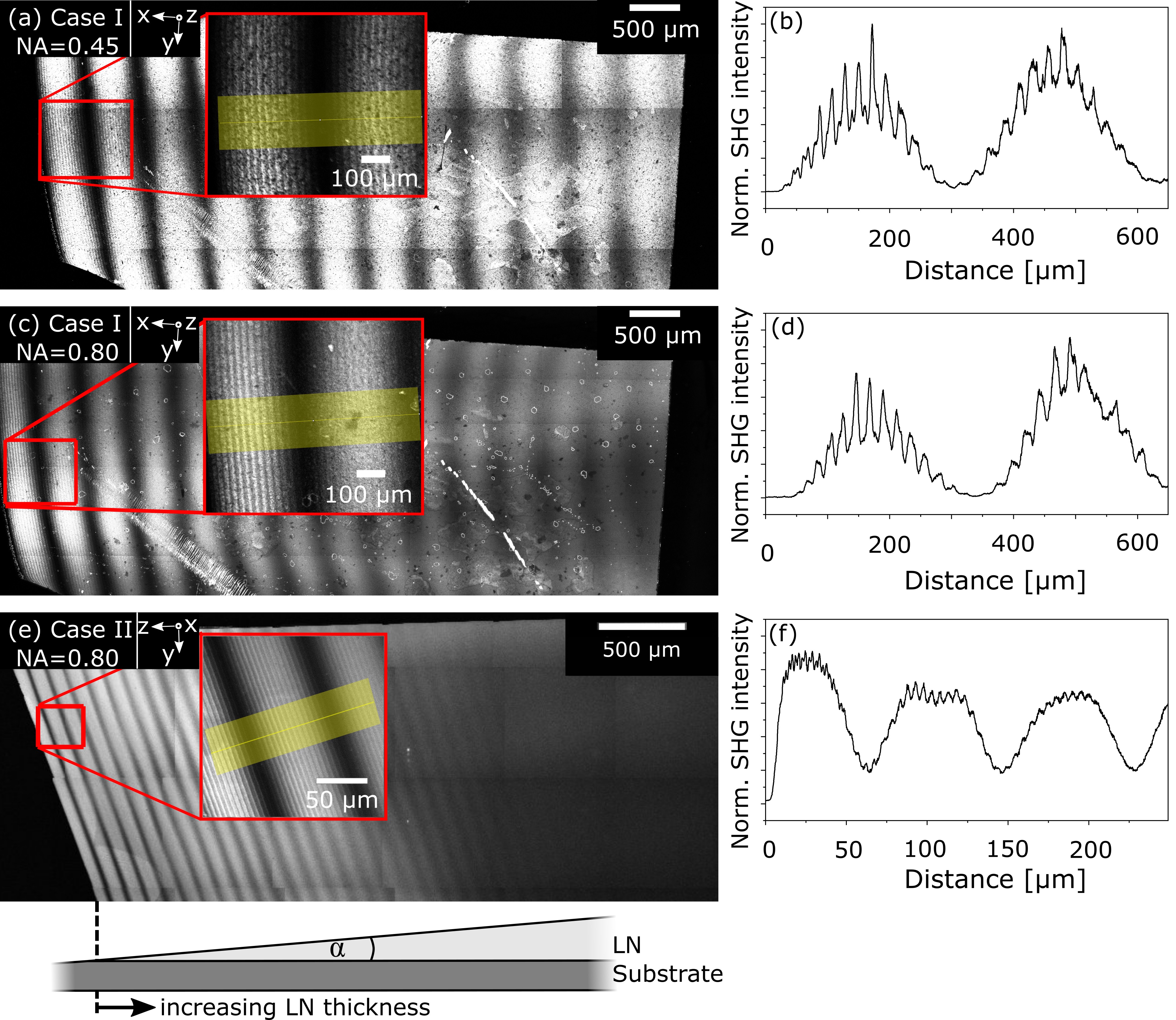}
		\caption{Large-scale images of the z- and x-cut wedges imaged under the conditions for (a) Case I, z-cut, NA~=~0.45, (c) Case I, z-cut NA~=~0.8, and (e) Case III, x-cut, NA~=~0.8. The insets present selected areas that were imaged with high resolution and used to extract the cross sections as displayed in panels (b), (d), and (f).}
		\label{fig 2}
	\end{figure*}
	
	To analyze the thickness-dependent coherent interaction length $l_{c}$ of the SHG signal we carefully investigated the above three cases I, II, and III. Figure \ref{fig 2}(a), (c), and (e) display large-scale  (stitched) SHG overview images taken at a 850~nm fundamental wavelength. As the complete sample cannot be focused at the same time for each tile a z-stack was recorded and projected onto the 2D plan. In all these images, the wedge thickness $t$ linearly increases from left to right, as denoted by the sketch at the bottom of the figure. Note that Fig. \ref{fig 2}(a) and (c) were recorded from the z-cut wedge with NA~=~0.45 and NA~=~0.80, respectively, while Fig. \ref{fig 2}(e) shows the SHG response across the x-cut wedge at a NA~=~0.80. In general, all images show characteristic oscillations in the SHG response, producing wavefronts that are aligned orthogonal to the thickness increase (i.e. the wedge slope). Notably, the periodicity measured for the long-wavelength oscillations is much shorter for the x-cut wedge ($\approx 70~\mu$m) as compared to the z-cut wedge ($\approx 250~\mu$m), a fact that solely stems from the roughly 4$\times$-larger wedge angle in case II, and not associated to different $l_{c}$-values.
	
	When analyzing these images in more detail, several more observations can be made. Firstly, we notice for the x-cut wedge in Fig. \ref{fig 2}(e) that oscillations are visible over more than a 2~mm width, with the contrast vanishing in the last third of the image to the right, only. A rough calculation allows to estimate the crystal thickness at that spot to measure approximately 70~$\mu$m, where reflections of co-propagating phase-matched light obviously influence the overall SH intensity. This value is more than 10$\times$ larger as compared to the axial optical resolution $\Delta z$ in LN, with the latter given to:
	\begin{equation}
		\Delta z = 1.772  \frac{n_{\omega} \lambda}{NA^2}   \approx 5 \mu\textnormal{m}.
	\end{equation}
			
	Secondly, apart from the long-period oscillation associated with the co-propagating phase matching process, short-range oscillations are visible. Most noticeably, they have wavelengths of approximately 20-25~$\mu$m and 5~$\mu$m for z- and x-cut LN wedges, respectively. Normalizing with respect to the inclination angles $\alpha_z$ and $\alpha_x$ yields thickness differences between neighboring intensity maxima in the range of 160-190~nm. These oscillations are associated with thin film interference effects of the fundamental and/or the SH signal, and also prominently appear in the numerical calculations as discussed below in further detail. \\
	As is apparent from these cross sections here, the thin film resonances significantly alter the overall SH signal intensity by a factor of 2 or more, over thickness variations that measure just a few ten nm. Even more dramatic, these oscillations are visible up to thicknesses of at least several ten $\mu$m. This means that in particular for quantitative measurements of SHG intensities, reflections and the co-propagating process need to be carefully evaluated even up to a 50~$\mu$m crystal thickness and beyond. Moreover, interface reflectivity, thin film interference, and the coherent interaction length $l_{c}$ all are governed by the dispersion of the refractive index. Hence, they will also be affected when inspecting crystals of a constant thickness while varying the incident wavelength. These findings are excellently supported through our numerical simulations and documented in the Supplement Information section. III.D.
	
	The third striking feature appears when analyzing the number of interference fringes, i.e. 12 and 15 oscillations, as counted in case I for the two different NAs used here, over the same crystal width [see Figs. \ref{fig 2}(a) and (c)]. When strictly interpreting every oscillation period as twice the coherent interaction length $l_{c}$ as governed by Eq. \ref{eqn:2}, the one and only sound conclusion is that  $l_{c}$ must depend on the NA of the objective lens in use; a larger NA thus must result in a shorter $l_c$, a fact that is absolutely not reflected by Eq. \ref{eqn:2}. Nevertheless, this - at first glance counter-intuitive - result can be readily explained when taking the nature of a focusing objective into account: \\
	 Eqs. \ref{eqn:1} and \ref{eqn:2} describe the phase matching condition for planar waves only, while focal beams always show a severe k-vector spread. In a simple picture this means that rays stemming from the outer rim within the focused beam, hit the sample to inspect under non-orthogonal conditions and hence take a longer effective trajectory when traveling through the crystal, as compared to central rays within the beam. Different parts of the focused beam, therefore, will be phasematched at different depths of the layer \cite{Rusing2019e}. A strongly focusing objective (large NA) thus will \textit{observe} an overall perceived shorter coherent interaction length, as parts of the beam will travel at a particularly larger angle through the material. 
	
	We have simulated this scenario for different NAs and display them in section III.C in the Supplemental Information. Note that a NA of $\approx$ 0.1 already very well reproduces the plane-wave condition as relevant to Eq. \ref{eqn:2}. As a conclusion, we thus recommend here to introduce the \emph{effective} coherent interaction length $l_{c, eff}$ when dealing with focused beams, rather than solely using $l_{c}$ as being valid strictly speaking for planar waves, only.
		
	\begin{figure*}[ht]
		\centering
		\includegraphics[width=0.95\linewidth]{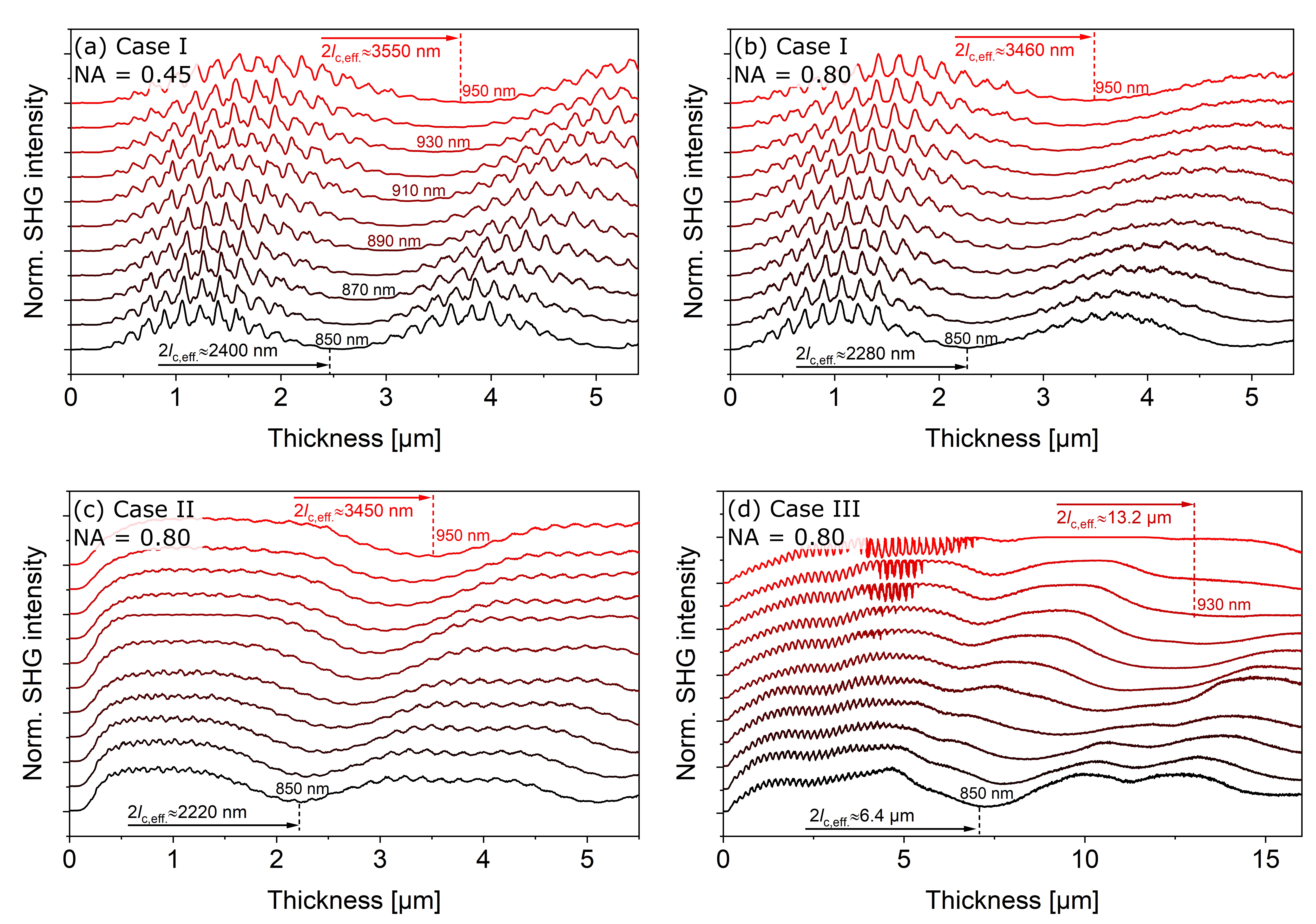}
		\caption{Measured SHG intensity profiles at a fundamental wavelength range from 850~nm to 950~nm for (a) Case I, NA~=~0.45, (b) Case I, NA~=~0.8, (c) Case II, NA~=~0.8, and (d) Case III, NA~=~0.8. In all datasets a systematic increase of the effective coherence length with wavelength can be observed due to dispersion.}
		\label{fig 3}
	\end{figure*}
	
	To analyze $l_{c, eff}$ for the three different cases as a function of wavelength and/or NA, we performed detailed scans at the thinner end of each wedge. To avoid any unambiguity or error through stitching or rendering, we used here single frames only that at least contained the first full oscillation, as shown by the insets in Fig. \ref{fig 2}(a), (c), and (e). Here, the focus was placed approximately at the center of the film at the thin edge of the wedge. As discussed in the supplement, the first oscillation period is least subjective to variations in focus positioning (See Supplemental Sec. IIIE). Moreover, lateral binning was applied in order to obtain the averaged intensity profiles, which is shown as the bold line in all these insets and displayed as cross sections in Figs. \ref{fig 2}(b), (d), and (e). \\
	For every case, the fundamental wavelength was swept from 850~nm to 950~nm via the microscope control software while the sample was kept at a constant location. Examples for as-obtained raw SHG data at the 850~nm fundamental wavelength, are shown in Figs. \ref{fig 2}(b), (d), and (f). For comparison reasons, all such SH intensity data were normalized with respect to the inclination angles $\alpha_z$ and $\alpha_x$, and then displayed in Fig. \ref{fig 3} as a function of thickness rather than lateral width. What becomes crucial, however, is to exactly define the origin in all these graphs, i.e. the exact position of zero thickness for each SHG profile. To do so, we first identified the location of the first significant SHG intensity ($> 0.1$~\%  of the maximum intensity). Then, profilometer scans across the wedge edge allowed us to quantify the residual step thickness at the thinner end of every LN wedge to approximately 50-100~nm. We therefore added to each position of significant first SHG intensity an offset of a 100~nm thickness, and conservatively assumed a confidence interval for the converted thicknesses of $\pm 100$~nm. Note, that the assumption of an initial step of 100~nm also fits well to our observations from numerical simulations, where the first local SHG maximum and hence the first significant SHG intensity as defined above, is typically observed for thicknesses of approximately 90~nm. This corresponds to the first maximum of the thin film interference of SHG light at 425~nm. A proof of that is seen from the excellent overlap of the measured and simulated data sets as illustrated later in Fig. \ref{fig 4}. It should be noted that the confidence interval of $\pm 100$~nm corresponds to less than 5\% of the total value of $2l_c$, which is always larger than 2.5~$\mu$m for all three cases I, II, and III. Therefore, any further errors in the measured data, e.g. imposed by the uncertainty in the wedge angles of $\pm 0.01$\textdegree, are negligible for wedge thicknesses of $< 20$~$\mu$m, which are at most considered for detailed quantitative analysis here.
	
	Figure \ref{fig 3} shows the obtained data sets as converted to the respective material thickness for all three cases. Each subfigure contains in total eleven datasets measured at fundamental wavelength from 850 - 950~nm in 10-nm increments. For optimal comparability, all datasets are normalized in their magnitude and displayed as a waterfall plot. Overall, for each case a distinct oscillation period is observed, that increases with increasing wavelength as expected from the decreasing dispersion. \\
	In detail for Case I, one can see an increase of the position of the first minimum from $2l_{c,eff} \approx 2400$~nm to  $2l_{c,eff} \approx 3550$~nm for the lower NA [Fig. \ref{fig 3}(a)], while a slightly shorter period is observed for the larger NA  [Fig. \ref{fig 3}(b)] with the position of the first minimum shifting from $2l_{c,eff} \approx 2280$~nm to  $2l_{c,eff} \approx 3460$~nm. In contrast, the corresponding periods for Case II [Fig. \ref{fig 3}(c)] are somewhat lower as compared to Case I (from $2l_{c,eff} \approx 2220$~nm to  $2l_{c,eff} \approx 3450$~nm for increasing wavelengths), although expected to be larger (see Tab. \ref{cases}). As discussed later and also clearly underlined by the numerical calculations, the influence of focusing on the effective coherent interaction length is less pronounced in Case II as compared to Case I, which is likely due to the slightly lower value of the overall refractive index of the extraordinary index resulting in less refraction, and hence a longer path length for the outer rays in a focused beam. \\
	Furthermore, we notice that the short-scale oscillation amplitudes related to the thin film interferences, are much lower for the x-cut sample (Cases II and III) as compared to the z-cut Case I. This is mostly due to the significantly different substrate reflectivity: The silicon substrate used to prepare the z-cut sample has a refractive index of $n = 3.63$ to $n=5.02$ at 850~nm to 425~nm wavelengths, respectively, while the epoxy features a moderate refractive index of $n \approx 1.52$ without any pronounced dispersion over the relevant wavelength range. This results in much larger reflections and hence thin film interference effects for the z-cut wedge used in Case I. \\
	Figure \ref{fig 3}(d) shows the data related to Case III. Compared to the other cases a more complex shape with more than one distinct minimum is observed. The dominating minimum, for example, observed at $2l_{c,eff} \approx 6400$~nm for 850~nm shows a much longer coherent interaction length compared to the cases discussed so far, and belongs to Case III as described in Tab. \ref{cases}. The position of this minimum shifts to higher wavelengths and can be identified up to $\lambda$ = 930~nm, where a length of $2l_{c,eff} \approx 13200$~nm is observed. Above that the contrast becomes too weak, while other processes may overlap with this effect. Apart from this minimum, a second set of oscillations is visible with minima observed, for example, at approximately 2.5~$\mu$m and 5~$\mu$m at 850~nm. They belong to a Case I process described by $P_y = d_{22} E_y^2$, and appear for this case here due to the unpolarized detection. The displayed data for Case III show that even when using unpolarized light, very complex SHG responses may be expected that need to be taken into account in every SHG experiment.

	\section{Discussion}
	
	\begin{figure}[ht]
		\centering
		\includegraphics[width=0.95\linewidth]{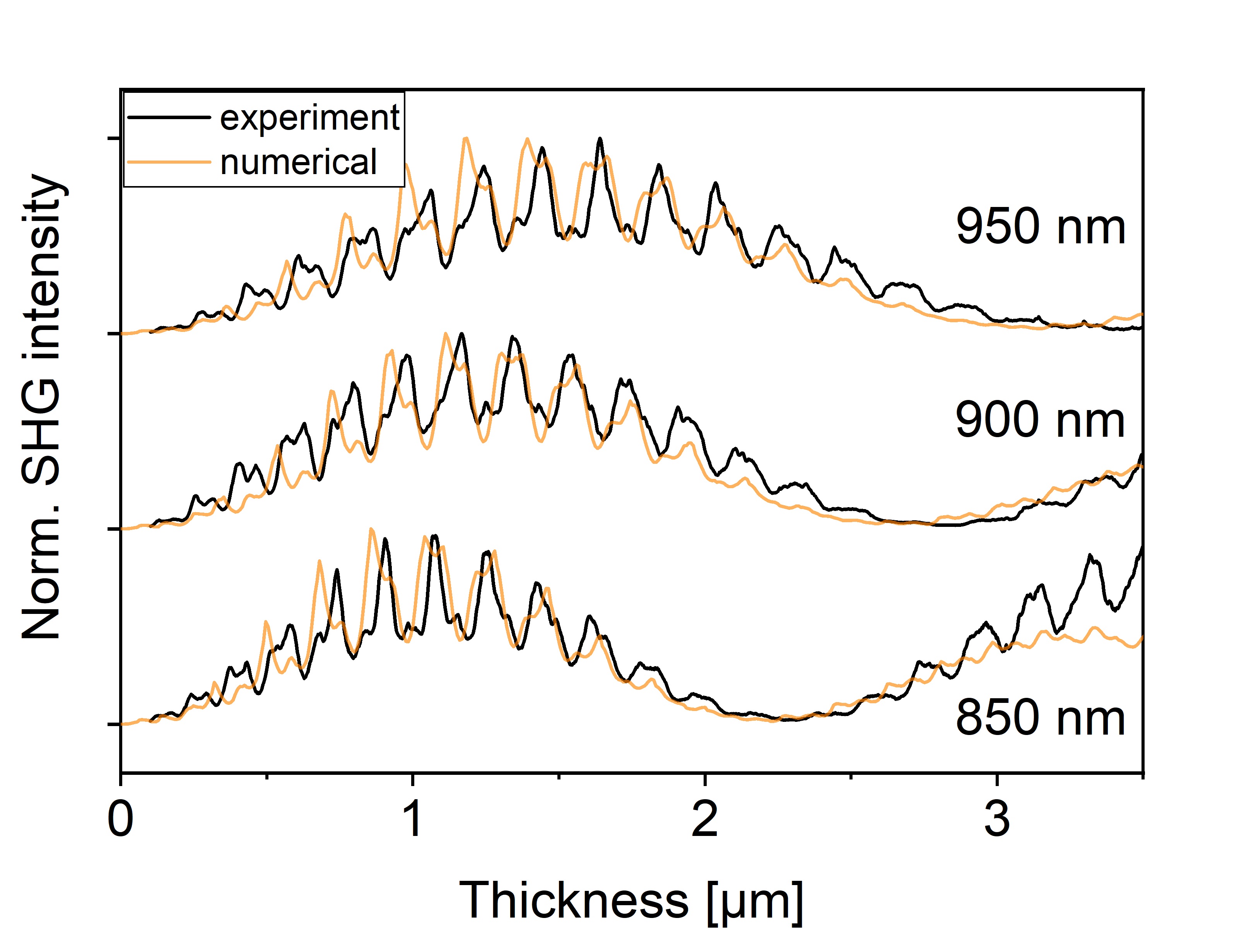}
		\caption{Exemplary comparison of calculated and measured intensity profiles for Case I, $NA$ = 0.8, at three different wavelengths of 850~nm, 900~nm, and 950~nm.}
		\label{fig 4}
	\end{figure}
	
	For quantitative comparison of the experimental data we have performed thorough numerical simulations. Figure \ref{fig 4} shows such an example comparing the numerically-calculated profiles to the experimental data of Case I with $NA$ = 0.8 at selected wavelengths. All datasets were normalized for maximum intensity and plotted with an artificial offset for better visibility. To allow for a detailed comparison of the overall shapes the SHG intensity was calculated in thickness increments of only 8~nm. Overall, the simulation reproduces the general shape, as well as the position, period, and magnitude of the short-scale and large-scale oscillations very well. These short-range oscillations belong to thin film interferences of either the fundamental and/or SHG beam as discussed in detail below. It should be noted that the experiment and simulated peaks show some slight deviations. For example at 950~nm, the double peaks visible around the overall maximum show a reversed intensity ratio in the experiment as compared to simulation. We believe that this results from the different reflectivity observed in experiment and simulation at the respective wavelength, due to the additional layer of glue present in the experiment, which is well backed up by numerical simulations as shown in the Supplemental Information section III.G. Furthermore, slight differences between the calculated refractive indices given via the Sellmeier equations \cite{Zelmon2008,Apnes1983} and the real refractive indices of the sample, might play an additional role. Moreover, our numerical calculations were always carried out with monochromatic beams, while the spectral width of pulse width effectively measures approximately 10~nm. However, we expect this to have a negligible impact on the simulated coherent length, because a larger wavelength will result in a longer coherent interaction length, while the opposite will be observed for the shorter wavelength of the pulse. Therefore, these effects are expected to cancel and only the contribution of the central wavelength will remain. In conclusion, this result demonstrates that the simulation is able not only to reproduce the large-scale oscillations originating from the coherent interaction length, but also various other features.
	
	As explained above for the experimental data it was considered that the thinnest, homogeneous thickness of the LN wedge is in the range of 50 to 100~nm based on mechanical profilometer scans. Therefore, to all experimental datasets the origin (0~nm thickness) was shifted by a constant thickness of 100~nm from the point of first significant SHG intensity, and this was accounted for in a respective thickness error bar ($\pm 100$~nm). This shift is also seen in Fig. \ref{fig 4}, where the experimental data are plotted from 100~nm onward only. This previously artificial offset yields a very good agreement with the numerical data, adding another independent argument to the plausibility of the minimum wedge thickness in the range of 50~nm to 100~nm, as well as the plausibility of the confidence interval.
	
	\begin{figure}
		\centering
		\includegraphics[width=0.95\linewidth]{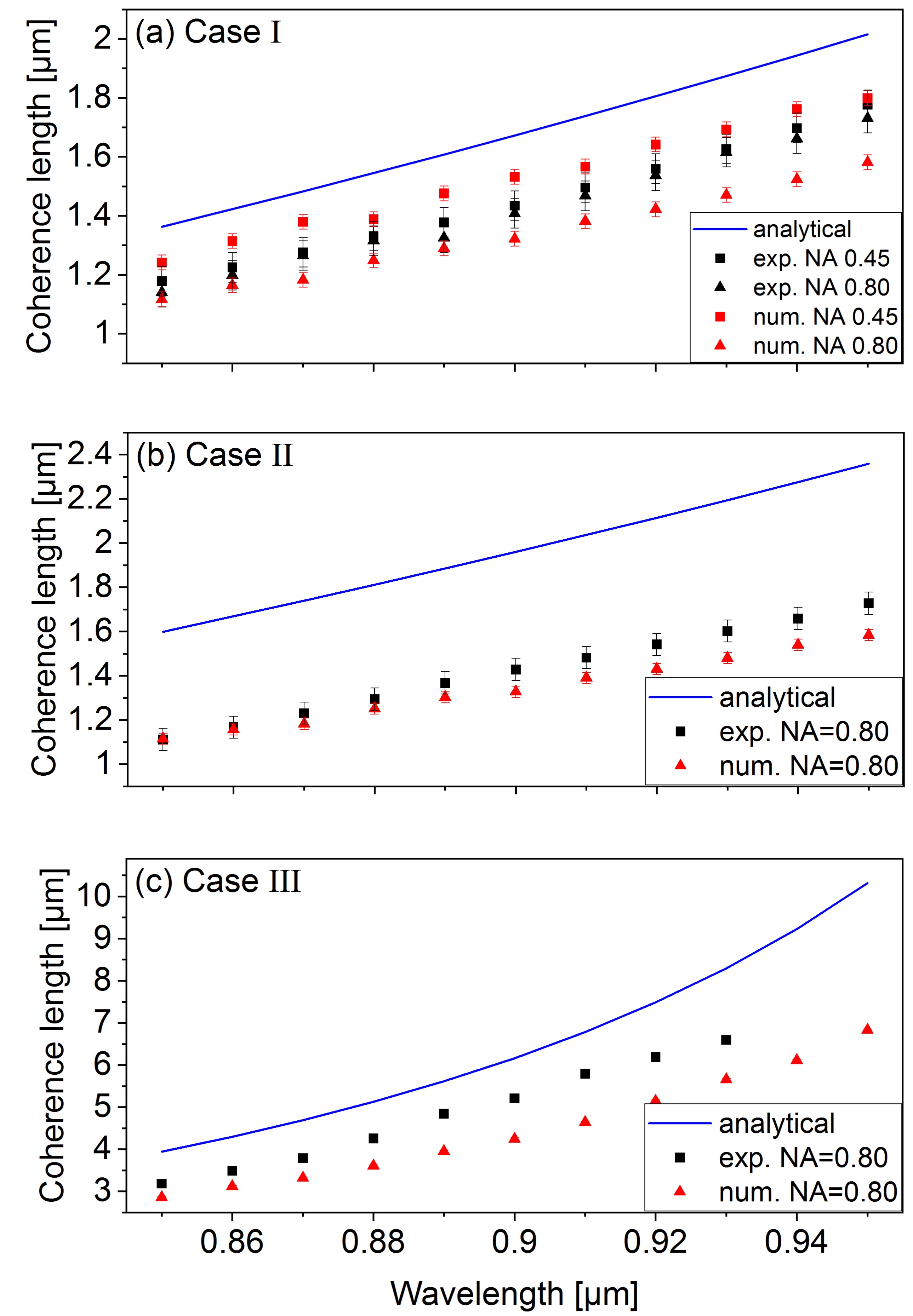}
		\caption{Experimentally and numerically-evaluated effective coherence interaction length $l_{c, eff}$ as a function of wavelenght for (a) Case I, (b) Case II, and (c) Case III.}
		\label{fig 5}
	\end{figure}
	
	In order to theoretically evaluate the coherent interaction length for all 3 cases, we have performed calculations as the one displayed in Fig. \ref{fig 4}. To save time, thickness increments of 100~nm (Cases I and II) and 200~nm (Case III) were used, however, using the same FEM grid for all calculations. This allows to analyze a much larger parameter space, as well as to test the influence of various different parameters as discussed in the Supplemental Information section III. Increments of 100~nm or 200~nm were deemed enough, as they allow to describe the coherent interaction length with a similar accuracy as in the experiment, as shown by the error bars in Fig. \ref{fig 5}.
	
	To compare the experimental and numerical data, we have determined the minima for all datasets. This data is plotted in Fig. \ref{fig 5} as a function of fundamental wavelength. Additionally, the graphs contain the predicted analytical behavior for plane wave incidence as describe by Eq. \ref{eqn:2} (continuous blue line). First, we immediately notice that the analytical result always yields a larger coherent interaction length as compared to the experiment and simulation. As discussed above, this is an effect of using focused beams, while Eq. \ref{eqn:2} is valid for plane waves, only. Indeed, the numerical data also display a shorter effective coherent length when compared to the analytical data, and fit well to our experimental results. In particular, the simulation is even able to reproduce the effect of a decreasing $l_{c, eff}$ with increasing $NA$, as displayed in Fig. \ref{fig 5}(a) for Case I. While the confidence intervals for the two $NA$ datasets in Fig. \ref{fig 5} do overlap, the experimentally deduced $l_{c, eff}$ for $NA$~=~0.8 is always larger as compared to simulation. Note, though that the effective coherent interaction length for $NA$~=~0.8 is only slightly smaller as compared to the $NA$~=~0.45 case, for all wavelengths.
	
	To exclude a systematic error, e.g. having chosen a wrong offset or different locations, we have repeated the Case I experiments for both NAs while rotating the sample by 90\textdegree. Then the detected SHG intensity is governed by $P_x = -d_{21} E_x^2$, however, leading to the same coherent interaction length and process as in Case I. As shown in the Supplemental Information section II, the determined $l_{c, eff}$ for both NAs and both orientations are the same, despite being independent measurements. Therefore, the slight deviations encountered here might stem from a sub-optimal illumination of the entry pupil of the NA~=~0.8 objective. Please note, that our simulation always considers an ideal planar wave back illumination of the objective. However, a lower illumination factor yields an effectively reduced numerical aperture, as shown in the Supplement. Hence, the observed effective coherent interaction length will be affected. 
	
	Nevertheless, the trends and slopes are very well reproduced. The excellent agreement between the behavior predicted by numerical calculation and the experiment confirms that indeed the co-propagating signal is detected in back-reflection geometry. It further shows that the numerical simulation is able to accurately reproduce many of the effects, and presents an important tool to improve the understanding of NLO microscopy experiments.
	
	\begin{figure}
		\centering
		\includegraphics[width=0.95\linewidth]{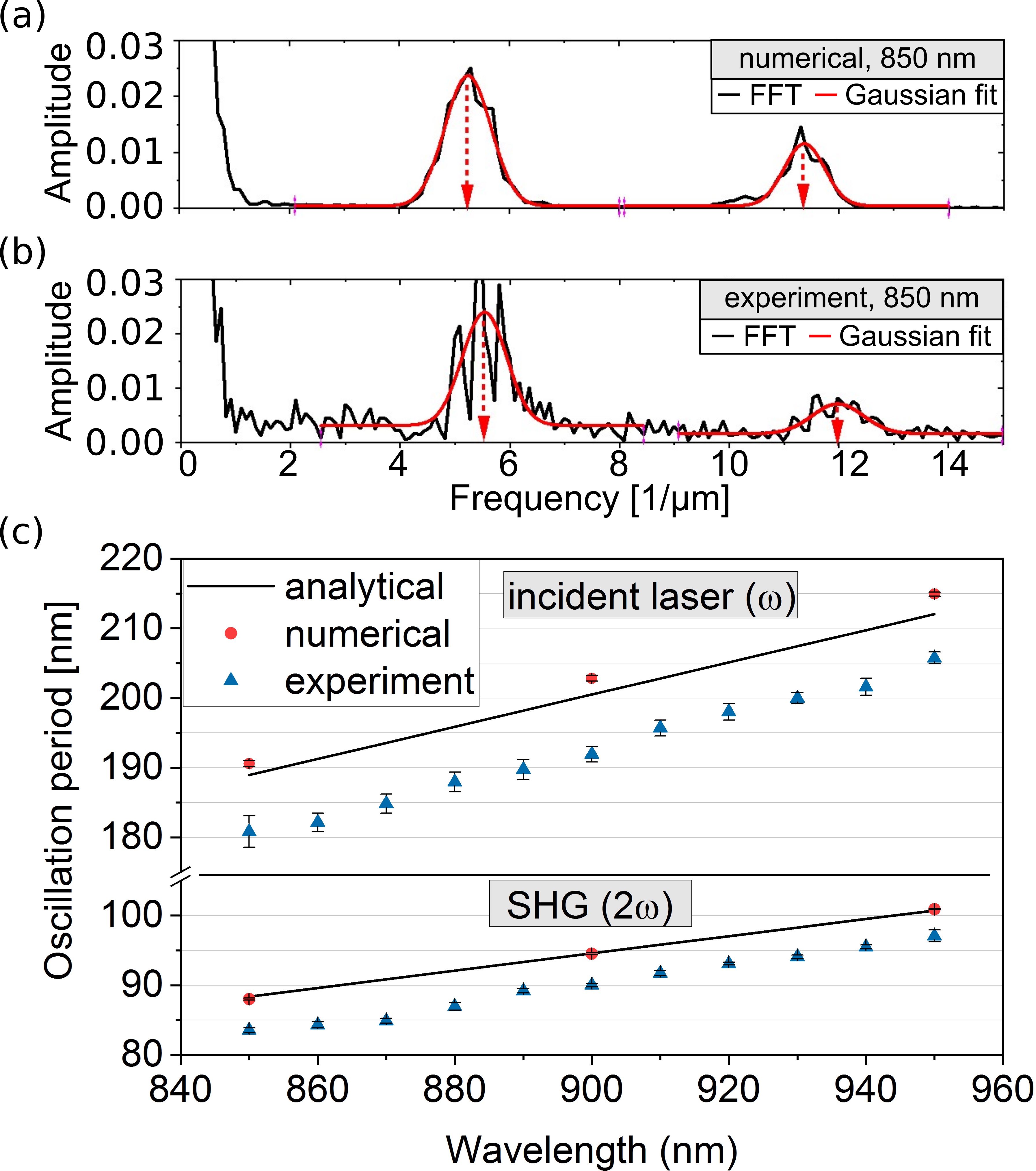}
		\caption{Subfigures (a) and (b) display an example for a FFT analysis of the datasets for Case I at an 850~nm fundamental wavelength and NA~=~0.8, which both show two distinct peaks identified as the interference length for the fundamental and SHG wavelength, respectively. (c) Based on the FFT analysis the oscillation periods for Case I were evaluated as a function of fundamental wavelength from the experimental data, as well as analytically and based on the high-resolution numerical data.}
		\label{fig_ripples}
	\end{figure}
	
	As mentioned above, the short-period oscillations present in most datasets in Figs. \ref{fig 3} and \ref{fig 4}, are due to thin film resonance effects of the fundamental and SHG beams. A resonant thickness $d$ or multiples $m$ of it, can be described by the well known relation
		\begin{equation}
			 d = m \frac{\lambda_{\omega}}{2n_{\omega}},
			 \label{resonance}
		\end{equation}
	where $\lambda_{\omega}$ is the wavelength at the fundamental or SHG frequency and $n_{\omega}$ its respective refractive index. To investigate the origin of these oscillations, we have performed a FFT analysis of the datasets for Case I [Fig. \ref{fig 3}(a)], which shows the strongest resonances, and the respective high resolution numerical data (Fig. \ref{fig 4}). The results for the 850 nm fundamental wavelength are displayed in Fig.~\ref{fig_ripples}(a) and (b), respectively. The FFT of both datasets, numerical and experimental, show two prominent peaks at spatial frequencies of approximately 5.5~$\mu$m$^{-1}$ and 11~$\mu$m$^{-1}$ belonging to periods of about 180~nm and 90~nm, respectively. Both fit exemplary well to the expected resonance periods for the fundamental and SHG light, respectively.
	
	Figure \ref{fig_ripples}(c) shows the extracted thin film resonance periods as a function of wavelength, together with the analytical calculation based on Eq.~\ref{resonance}. In contrast to the analysis of the effective coherent length, here the numerical aperture plays no prominent role for the thin film resonances period. Moreover, the numerical and analytical calculations align well and yield similar results. This is expected as the oscillations (a) are aligned parallel to the interface planes, as seen in the focal field plots in the supplement, and (b) the refractive indices for both, the analytical and numerical calculations, are based on the same refractive index \cite{Zelmon2008}. The experimental data shows the same slope and behavior as do the numerical and analytical data, however, is systematically shifted to slightly smaller periods for both, the SHG and fundamental wavelength. A potential explanation is a slightly different refractive index of the used 5\% MgO-doped LN as compared to literature values. Please note, a difference in the \emph{magnitude} of the refractive index will not influence the calculated coherent length, as this is dependent on dispersion, i.e. differences between the refractive index at the fundamental and SHG wavelength, which are similar even when the absolute refractive index is shifted due to variations in ambient conditions, doping, defects or fabrication \cite{Volk2008,Wong2002}. Nevertheless, the excellent agreement of the FFT data (a),(b) and the slopes in (c) demonstrate the capability of the simulation to also describe resonance effects.
	

	\section{Conclusion}
	\label{Sec_Conclusion}

In this work we have investigated the impact of confinement and reflective interfaces on the SHG signal in standard back reflection microscopy. In such a setup it is commonly assumed that the detected SHG intensity exclusively reflects the counter-propagating phase-matched SHG signal, generally stemming from a tiny effective interaction volume only, due to the short coherent interaction length $l_c$. In addition, that phase mismatch $\Delta k$, hence, must always be positive independent of wavelength, which has severe implications on the expected SHG signatures as shown by Kaneshiro et al. \cite{Kaneshiro2008}. 

However, whenever reflective interfaces of a few percent reflectivity and/or confining (nano)structures are involved, which thin films are \textit{per se}, also the co-propagating phase-matched signal is reflect back into the detector, and superposes to the counter-propagating signal. Even more dramatically, $l_c$ for the co-propagating signal reaches significantly larger values as compared to the counter-propagating case. Additionally for confined structures, significant enhancement of the SHG response can be observed due to thin-film interference effects of either the fundamental or SHG beam. 

As our experiments and numerical simulations clearly show, these confinement effects are not just limited to the film thickness on the order of the wavelength, as it is often assumed for linear microscopy, but play a major role even up to macroscopic thicknesses of a few ten micrometers. Therefore, from the viewpoint of quantitative SHG microscopy, a crystal may only be considered \emph{bulk} whenever its thickness measures at least several times the axial resolution, or when spurious reflections are significantly suppressed, e.g. by index matching oil. Our results are particularly relevant when obtaining quantitative SHG values, not only of samples with varying thicknesses, but also when tuning over a broader wavelength range, solely due to dispersive and polarization effects. This can be of particular relevance when examining for instance ferroelectric thin films with thicknesses ranging from a few nm up to several $\mu$m, where SHG microscopy and polarimetry is widely used to analyze local bulk and domain wall symmetries \cite{Cherifi-Hertel2021,Denev2011,Zhao2020,Zhao2019a}. 

In this work, we have considered the SHG effect as an archetypal nonlinear optical (NLO) effect. While phase-matching length and wavelengths will depend on the specific NLO effect, it can be expected that similar results will influence the NLO response in other processes as well, e.g. CARS or THG. To judge on the specific influence and magnitude of confinement and reflections in those cases, further theoretical and experimental studies are needed. 

In conclusion, our work demonstrates that reflections can play a crucial part in NLO experiments and should not be neglected. This improved understanding of NLO microscopy, hence, will lead to a much more detailed experimental interpretation and - as already suggested for SHG on TFLN \cite{Rusing2019e,Sandkuijl2013} - potential super-resolution applications of NLO microscopy.

	\section*{Supplemental Material}
	\label{sec_supple}
Additional details on the experimental setup and the sample preparation, additional experiments and a numerical discussion on the influence of various parameters can be found in the Supplemental file.
	
		\section*{Data availability}
	The data that support the findings of this study are available from the corresponding author upon reasonable request.
	
	\begin{acknowledgments}
		We express our gratitude for financial support by the Deutsche Forschungsgemeinschaft (DFG) through joint DFG-ANR project „TOPELEC“ (EN 434/41-1 and ANR-18-CE92-0052-1), the CRC1415 (ID: 417590517) the FOR5044 (ID: 426703838; \url{https://www.for5044.de}) as well as by the Würzburg-Dresden Cluster of Excellence on “Complexity and Topology in Quantum Matter” - ct.qmat (EXC 2147; ID: 39085490). We also acknowledge the excellent support by the Light Microscopy Facility, a Core Facility of the CMCB Technology Platform at TU Dresden, where the SHG analysis was performed. Further, we thank D. Bieberstein and T. Gemming from IFW Dresden for assistance with dicing of the wafers, P. Mackwitz for providing help with Laser Scanning Microscopy and SGS Fresenius GmbH for fabrication of the wedge samples.
	\end{acknowledgments}


\begin{thebibliography}{42}%
	\makeatletter
	\providecommand \@ifxundefined [1]{%
		\@ifx{#1\undefined}
	}%
	\providecommand \@ifnum [1]{%
		\ifnum #1\expandafter \@firstoftwo
		\else \expandafter \@secondoftwo
		\fi
	}%
	\providecommand \@ifx [1]{%
		\ifx #1\expandafter \@firstoftwo
		\else \expandafter \@secondoftwo
		\fi
	}%
	\providecommand \natexlab [1]{#1}%
	\providecommand \enquote  [1]{``#1''}%
	\providecommand \bibnamefont  [1]{#1}%
	\providecommand \bibfnamefont [1]{#1}%
	\providecommand \citenamefont [1]{#1}%
	\providecommand \href@noop [0]{\@secondoftwo}%
	\providecommand \href [0]{\begingroup \@sanitize@url \@href}%
	\providecommand \@href[1]{\@@startlink{#1}\@@href}%
	\providecommand \@@href[1]{\endgroup#1\@@endlink}%
	\providecommand \@sanitize@url [0]{\catcode `\\12\catcode `\$12\catcode
		`\&12\catcode `\#12\catcode `\^12\catcode `\_12\catcode `\%12\relax}%
	\providecommand \@@startlink[1]{}%
	\providecommand \@@endlink[0]{}%
	\providecommand \url  [0]{\begingroup\@sanitize@url \@url }%
	\providecommand \@url [1]{\endgroup\@href {#1}{\urlprefix }}%
	\providecommand \urlprefix  [0]{URL }%
	\providecommand \Eprint [0]{\href }%
	\providecommand \doibase [0]{https://doi.org/}%
	\providecommand \selectlanguage [0]{\@gobble}%
	\providecommand \bibinfo  [0]{\@secondoftwo}%
	\providecommand \bibfield  [0]{\@secondoftwo}%
	\providecommand \translation [1]{[#1]}%
	\providecommand \BibitemOpen [0]{}%
	\providecommand \bibitemStop [0]{}%
	\providecommand \bibitemNoStop [0]{.\EOS\space}%
	\providecommand \EOS [0]{\spacefactor3000\relax}%
	\providecommand \BibitemShut  [1]{\csname bibitem#1\endcsname}%
	\let\auto@bib@innerbib\@empty
	\bibitem [{\citenamefont {Squier}\ and\ \citenamefont
		{M{\"{u}}ller}(2001)}]{Squier2001}%
	\BibitemOpen
	\bibfield  {author} {\bibinfo {author} {\bibfnamefont {J.}~\bibnamefont
			{Squier}}\ and\ \bibinfo {author} {\bibfnamefont {M.}~\bibnamefont
			{M{\"{u}}ller}},\ }\bibfield  {title} {\enquote {\bibinfo {title} {{High
					resolution nonlinear microscopy: A review of sources and methods for
					achieving optimal imaging}},}\ }\href {https://doi.org/10.1063/1.1379598}
	{\bibfield  {journal} {\bibinfo  {journal} {Review of Scientific
				Instruments}\ }\textbf {\bibinfo {volume} {72}},\ \bibinfo {pages}
		{2855--2867} (\bibinfo {year} {2001})}\BibitemShut {NoStop}%
	\bibitem [{\citenamefont {Cox}(2011)}]{Cox2011}%
	\BibitemOpen
	\bibfield  {author} {\bibinfo {author} {\bibfnamefont {G.}~\bibnamefont
			{Cox}},\ }\bibfield  {title} {\enquote {\bibinfo {title} {{Biological
					applications of second harmonic imaging}},}\ }\href
	{https://doi.org/10.1007/s12551-011-0052-9} {\bibfield  {journal} {\bibinfo
			{journal} {Biophysical Reviews}\ }\textbf {\bibinfo {volume} {3}},\ \bibinfo
		{pages} {131--141} (\bibinfo {year} {2011})}\BibitemShut {NoStop}%
	\bibitem [{\citenamefont {Kumar}\ \emph {et~al.}(2013)\citenamefont {Kumar},
		\citenamefont {Najmaei}, \citenamefont {Cui}, \citenamefont {Ceballos},
		\citenamefont {Ajayan}, \citenamefont {Lou},\ and\ \citenamefont
		{Zhao}}]{Kumar2013}%
	\BibitemOpen
	\bibfield  {author} {\bibinfo {author} {\bibfnamefont {N.}~\bibnamefont
			{Kumar}}, \bibinfo {author} {\bibfnamefont {S.}~\bibnamefont {Najmaei}},
		\bibinfo {author} {\bibfnamefont {Q.}~\bibnamefont {Cui}}, \bibinfo {author}
		{\bibfnamefont {F.}~\bibnamefont {Ceballos}}, \bibinfo {author}
		{\bibfnamefont {P.~M.}\ \bibnamefont {Ajayan}}, \bibinfo {author}
		{\bibfnamefont {J.}~\bibnamefont {Lou}},\ and\ \bibinfo {author}
		{\bibfnamefont {H.}~\bibnamefont {Zhao}},\ }\bibfield  {title} {\enquote
		{\bibinfo {title} {{Second harmonic microscopy of monolayer MoS2}},}\ }\href
	{https://doi.org/10.1103/PhysRevB.87.161403} {\bibfield  {journal} {\bibinfo
			{journal} {Physical Review B}\ }\textbf {\bibinfo {volume} {87}},\ \bibinfo
		{pages} {161403} (\bibinfo {year} {2013})}\ \BibitemShut {NoStop}%
	\bibitem [{\citenamefont {Carriles}\ \emph {et~al.}(2009)\citenamefont
		{Carriles}, \citenamefont {Schafer}, \citenamefont {Sheetz}, \citenamefont
		{Field}, \citenamefont {Cisek}, \citenamefont {Barzda}, \citenamefont
		{Sylvester},\ and\ \citenamefont {Squier}}]{Carriles2009}%
	\BibitemOpen
	\bibfield  {author} {\bibinfo {author} {\bibfnamefont {R.}~\bibnamefont
			{Carriles}}, \bibinfo {author} {\bibfnamefont {D.~N.}\ \bibnamefont
			{Schafer}}, \bibinfo {author} {\bibfnamefont {K.~E.}\ \bibnamefont {Sheetz}},
		\bibinfo {author} {\bibfnamefont {J.~J.}\ \bibnamefont {Field}}, \bibinfo
		{author} {\bibfnamefont {R.}~\bibnamefont {Cisek}}, \bibinfo {author}
		{\bibfnamefont {V.}~\bibnamefont {Barzda}}, \bibinfo {author} {\bibfnamefont
			{A.~W.}\ \bibnamefont {Sylvester}},\ and\ \bibinfo {author} {\bibfnamefont
			{J.~A.}\ \bibnamefont {Squier}},\ }\bibfield  {title} {\enquote {\bibinfo
			{title} {{Invited Review Article: Imaging techniques for harmonic and
					multiphoton absorption fluorescence microscopy}},}\ }\href
	{https://doi.org/10.1063/1.3184828} {\bibfield  {journal} {\bibinfo
			{journal} {Review of Scientific Instruments}\ }\textbf {\bibinfo {volume}
			{80}} (\bibinfo {year} {2009})}\BibitemShut {NoStop}%
	\bibitem [{\citenamefont {Oron}\ \emph {et~al.}(2004)\citenamefont {Oron},
		\citenamefont {Yelin}, \citenamefont {Tal}, \citenamefont {Raz},
		\citenamefont {Fachima},\ and\ \citenamefont {Silberberg}}]{Oron2004}%
	\BibitemOpen
	\bibfield  {author} {\bibinfo {author} {\bibfnamefont {D.}~\bibnamefont
			{Oron}}, \bibinfo {author} {\bibfnamefont {D.}~\bibnamefont {Yelin}},
		\bibinfo {author} {\bibfnamefont {E.}~\bibnamefont {Tal}}, \bibinfo {author}
		{\bibfnamefont {S.}~\bibnamefont {Raz}}, \bibinfo {author} {\bibfnamefont
			{R.}~\bibnamefont {Fachima}},\ and\ \bibinfo {author} {\bibfnamefont
			{Y.}~\bibnamefont {Silberberg}},\ }\bibfield  {title} {\enquote {\bibinfo
			{title} {{Depth-resolved structural imaging by third-harmonic generation
					microscopy}},}\ }\href {https://doi.org/10.1016/S1047-8477(03)00125-4}
	{\bibfield  {journal} {\bibinfo  {journal} {Journal of Structural Biology}\
		}\textbf {\bibinfo {volume} {147}},\ \bibinfo {pages} {3--11} (\bibinfo
		{year} {2004})}\BibitemShut {NoStop}%
	\bibitem [{\citenamefont {Golovan}\ \emph {et~al.}(2012)\citenamefont
		{Golovan}, \citenamefont {Gonchar}, \citenamefont {Osminkina}, \citenamefont
		{Timoshenko}, \citenamefont {Petrov},\ and\ \citenamefont
		{Yakovlev}}]{Golovan2012}%
	\BibitemOpen
	\bibfield  {author} {\bibinfo {author} {\bibfnamefont {L.}~\bibnamefont
			{Golovan}}, \bibinfo {author} {\bibfnamefont {K.}~\bibnamefont {Gonchar}},
		\bibinfo {author} {\bibfnamefont {L.}~\bibnamefont {Osminkina}}, \bibinfo
		{author} {\bibfnamefont {V.}~\bibnamefont {Timoshenko}}, \bibinfo {author}
		{\bibfnamefont {G.}~\bibnamefont {Petrov}},\ and\ \bibinfo {author}
		{\bibfnamefont {V.}~\bibnamefont {Yakovlev}},\ }\bibfield  {title} {\enquote
		{\bibinfo {title} {{Coherent anti-Stokes Raman scattering in silicon nanowire
					ensembles}},}\ }\href {https://doi.org/10.1002/lapl.201110099} {\bibfield
		{journal} {\bibinfo  {journal} {Laser Physics Letters}\ }\textbf {\bibinfo
			{volume} {9}},\ \bibinfo {pages} {145--150} (\bibinfo {year}
		{2012})}\BibitemShut {NoStop}%
	\bibitem [{\citenamefont {Day}\ \emph {et~al.}(2011)\citenamefont {Day},
		\citenamefont {Domke}, \citenamefont {Rago}, \citenamefont {Kano},
		\citenamefont {Hamaguchi}, \citenamefont {Vartiainen},\ and\ \citenamefont
		{Bonn}}]{Day2011}%
	\BibitemOpen
	\bibfield  {author} {\bibinfo {author} {\bibfnamefont {J.~P.~R.}\
			\bibnamefont {Day}}, \bibinfo {author} {\bibfnamefont {K.~F.}\ \bibnamefont
			{Domke}}, \bibinfo {author} {\bibfnamefont {G.}~\bibnamefont {Rago}},
		\bibinfo {author} {\bibfnamefont {H.}~\bibnamefont {Kano}}, \bibinfo {author}
		{\bibfnamefont {H.-O.}\ \bibnamefont {Hamaguchi}}, \bibinfo {author}
		{\bibfnamefont {E.~M.}\ \bibnamefont {Vartiainen}},\ and\ \bibinfo {author}
		{\bibfnamefont {M.}~\bibnamefont {Bonn}},\ }\bibfield  {title} {\enquote
		{\bibinfo {title} {{Quantitative Coherent Anti-Stokes Raman Scattering (CARS)
					Microscopy}},}\ }\href {https://doi.org/10.1021/jp200606e} {\bibfield
		{journal} {\bibinfo  {journal} {The Journal of Physical Chemistry B}\
		}\textbf {\bibinfo {volume} {115}},\ \bibinfo {pages} {7713--7725} (\bibinfo
		{year} {2011})}\BibitemShut {NoStop}%
	\bibitem [{\citenamefont {Djaker}\ \emph {et~al.}(2007)\citenamefont {Djaker},
		\citenamefont {Lenne}, \citenamefont {Marguet}, \citenamefont {Colonna},
		\citenamefont {Hadjur},\ and\ \citenamefont {Rigneault}}]{Djaker2007}%
	\BibitemOpen
	\bibfield  {author} {\bibinfo {author} {\bibfnamefont {N.}~\bibnamefont
			{Djaker}}, \bibinfo {author} {\bibfnamefont {P.-F.}\ \bibnamefont {Lenne}},
		\bibinfo {author} {\bibfnamefont {D.}~\bibnamefont {Marguet}}, \bibinfo
		{author} {\bibfnamefont {A.}~\bibnamefont {Colonna}}, \bibinfo {author}
		{\bibfnamefont {C.}~\bibnamefont {Hadjur}},\ and\ \bibinfo {author}
		{\bibfnamefont {H.}~\bibnamefont {Rigneault}},\ }\bibfield  {title} {\enquote
		{\bibinfo {title} {{Coherent anti-Stokes Raman scattering microscopy (CARS):
					Instrumentation and applications}},}\ }\href
	{https://doi.org/10.1016/j.nima.2006.10.057} {\bibfield  {journal} {\bibinfo
			{journal} {Nuclear Instruments and Methods in Physics Research Section A:
				Accelerators, Spectrometers, Detectors and Associated Equipment}\ }\textbf
		{\bibinfo {volume} {571}},\ \bibinfo {pages} {177--181} (\bibinfo {year}
		{2007})}\BibitemShut {NoStop}%
	\bibitem [{\citenamefont {Cherifi-Hertel}\ \emph {et~al.}(2017)\citenamefont
		{Cherifi-Hertel}, \citenamefont {Bulou}, \citenamefont {Hertel},
		\citenamefont {Taupier}, \citenamefont {Dorkenoo}, \citenamefont {Andreas},
		\citenamefont {Guyonnet}, \citenamefont {Gaponenko}, \citenamefont {Gallo},\
		and\ \citenamefont {Paruch}}]{Cherifi-Hertel2017}%
	\BibitemOpen
	\bibfield  {author} {\bibinfo {author} {\bibfnamefont {S.}~\bibnamefont
			{Cherifi-Hertel}}, \bibinfo {author} {\bibfnamefont {H.}~\bibnamefont
			{Bulou}}, \bibinfo {author} {\bibfnamefont {R.}~\bibnamefont {Hertel}},
		\bibinfo {author} {\bibfnamefont {G.}~\bibnamefont {Taupier}}, \bibinfo
		{author} {\bibfnamefont {K.~D.~H.}\ \bibnamefont {Dorkenoo}}, \bibinfo
		{author} {\bibfnamefont {C.}~\bibnamefont {Andreas}}, \bibinfo {author}
		{\bibfnamefont {J.}~\bibnamefont {Guyonnet}}, \bibinfo {author}
		{\bibfnamefont {I.}~\bibnamefont {Gaponenko}}, \bibinfo {author}
		{\bibfnamefont {K.}~\bibnamefont {Gallo}},\ and\ \bibinfo {author}
		{\bibfnamefont {P.}~\bibnamefont {Paruch}},\ }\bibfield  {title} {\enquote
		{\bibinfo {title} {{Non-Ising and chiral ferroelectric domain walls revealed
					by nonlinear optical microscopy}},}\ }\href
	{https://doi.org/10.1038/ncomms15768} {\bibfield  {journal} {\bibinfo
			{journal} {Nature Communications}\ }\textbf {\bibinfo {volume} {8}},\
		\bibinfo {pages} {15768} (\bibinfo {year} {2017})}\BibitemShut {NoStop}%
	\bibitem [{\citenamefont {Cherifi-Hertel}\ \emph {et~al.}(2021)\citenamefont
		{Cherifi-Hertel}, \citenamefont {Voulot}, \citenamefont {Acevedo-Salas},
		\citenamefont {Zhang}, \citenamefont {Cr{\'{e}}gut}, \citenamefont
		{Dorkenoo},\ and\ \citenamefont {Hertel}}]{Cherifi-Hertel2021}%
	\BibitemOpen
	\bibfield  {author} {\bibinfo {author} {\bibfnamefont {S.}~\bibnamefont
			{Cherifi-Hertel}}, \bibinfo {author} {\bibfnamefont {C.}~\bibnamefont
			{Voulot}}, \bibinfo {author} {\bibfnamefont {U.}~\bibnamefont
			{Acevedo-Salas}}, \bibinfo {author} {\bibfnamefont {Y.}~\bibnamefont
			{Zhang}}, \bibinfo {author} {\bibfnamefont {O.}~\bibnamefont {Cr{\'{e}}gut}},
		\bibinfo {author} {\bibfnamefont {K.~D.}\ \bibnamefont {Dorkenoo}},\ and\
		\bibinfo {author} {\bibfnamefont {R.}~\bibnamefont {Hertel}},\ }\bibfield
	{title} {\enquote {\bibinfo {title} {{Shedding light on non-Ising polar
					domain walls: Insight from second harmonic generation microscopy and
					polarimetry analysis}},}\ }\href {https://doi.org/10.1063/5.0037286}
	{\bibfield  {journal} {\bibinfo  {journal} {Journal of Applied Physics}\
		}\textbf {\bibinfo {volume} {129}},\ \bibinfo {pages} {081101} (\bibinfo
		{year} {2021})}\BibitemShut {NoStop}%
	\bibitem [{\citenamefont {R{\"{u}}sing}, \citenamefont {Zhao},\ and\
		\citenamefont {Mookherjea}(2019)}]{Rusing2019e}%
	\BibitemOpen
	\bibfield  {author} {\bibinfo {author} {\bibfnamefont {M.}~\bibnamefont
			{R{\"{u}}sing}}, \bibinfo {author} {\bibfnamefont {J.}~\bibnamefont {Zhao}},\
		and\ \bibinfo {author} {\bibfnamefont {S.}~\bibnamefont {Mookherjea}},\
	}\bibfield  {title} {\enquote {\bibinfo {title} {{Second harmonic microscopy
					of poled x-cut thin film lithium niobate: Understanding the contrast
					mechanism}},}\ }\href {https://doi.org/10.1063/1.5113727} {\bibfield
		{journal} {\bibinfo  {journal} {Journal of Applied Physics}\ }\textbf
		{\bibinfo {volume} {126}},\ \bibinfo {pages} {114105} (\bibinfo {year}
		{2019})} \BibitemShut {NoStop}%
	\bibitem [{\citenamefont {Spychala}\ \emph {et~al.}(2017)\citenamefont
		{Spychala}, \citenamefont {Berth}, \citenamefont {Widhalm}, \citenamefont
		{R{\"{u}}sing}, \citenamefont {Wang}, \citenamefont {Sanna},\ and\
		\citenamefont {Zrenner}}]{Spychala2017}%
	\BibitemOpen
	\bibfield  {author} {\bibinfo {author} {\bibfnamefont {K.~J.}\ \bibnamefont
			{Spychala}}, \bibinfo {author} {\bibfnamefont {G.}~\bibnamefont {Berth}},
		\bibinfo {author} {\bibfnamefont {A.}~\bibnamefont {Widhalm}}, \bibinfo
		{author} {\bibfnamefont {M.}~\bibnamefont {R{\"{u}}sing}}, \bibinfo {author}
		{\bibfnamefont {L.}~\bibnamefont {Wang}}, \bibinfo {author} {\bibfnamefont
			{S.}~\bibnamefont {Sanna}},\ and\ \bibinfo {author} {\bibfnamefont
			{A.}~\bibnamefont {Zrenner}},\ }\bibfield  {title} {\enquote {\bibinfo
			{title} {{Impact of carbon-ion implantation on the nonlinear optical
					susceptibility of LiNbO$_3$}},}\ }\href {https://doi.org/10.1364/OE.25.021444}
	{\bibfield  {journal} {\bibinfo  {journal} {Optics Express}\ }\textbf
		{\bibinfo {volume} {25}},\ \bibinfo {pages} {21444} (\bibinfo {year}
		{2017})}\BibitemShut {NoStop}%
	\bibitem [{\citenamefont {Reitzig}\ \emph {et~al.}(2021)\citenamefont
		{Reitzig}, \citenamefont {R{\"{u}}sing}, \citenamefont {Zhao}, \citenamefont
		{Kirbus}, \citenamefont {Mookherjea},\ and\ \citenamefont
		{Eng}}]{Reitzig2021}%
	\BibitemOpen
	\bibfield  {author} {\bibinfo {author} {\bibfnamefont {S.}~\bibnamefont
			{Reitzig}}, \bibinfo {author} {\bibfnamefont {M.}~\bibnamefont
			{R{\"{u}}sing}}, \bibinfo {author} {\bibfnamefont {J.}~\bibnamefont {Zhao}},
		\bibinfo {author} {\bibfnamefont {B.}~\bibnamefont {Kirbus}}, \bibinfo
		{author} {\bibfnamefont {S.}~\bibnamefont {Mookherjea}},\ and\ \bibinfo
		{author} {\bibfnamefont {L.~M.}\ \bibnamefont {Eng}},\ }\bibfield  {title}
	{\enquote {\bibinfo {title} {{“Seeing Is Believing”—In-Depth Analysis
					by Co-Imaging of Periodically-Poled X-Cut Lithium Niobate Thin Films}},}\
	}\href {https://doi.org/10.3390/cryst11030288} {\bibfield  {journal}
		{\bibinfo  {journal} {Crystals}\ }\textbf {\bibinfo {volume} {11}},\ \bibinfo
		{pages} {288} (\bibinfo {year} {2021})}\BibitemShut {NoStop}%
	\bibitem [{\citenamefont {Berth}\ \emph {et~al.}(2007)\citenamefont {Berth},
		\citenamefont {Quiring}, \citenamefont {Sohler},\ and\ \citenamefont
		{Zrenner}}]{Berth2007}%
	\BibitemOpen
	\bibfield  {author} {\bibinfo {author} {\bibfnamefont {G.}~\bibnamefont
			{Berth}}, \bibinfo {author} {\bibfnamefont {V.}~\bibnamefont {Quiring}},
		\bibinfo {author} {\bibfnamefont {W.}~\bibnamefont {Sohler}},\ and\ \bibinfo
		{author} {\bibfnamefont {A.}~\bibnamefont {Zrenner}},\ }\bibfield  {title}
	{\enquote {\bibinfo {title} {{Depth-Resolved Analysis of Ferroelectric Domain
					Structures in Ti:PPLN Waveguides by Nonlinear Confocal Laser Scanning
					Microscopy}},}\ }\href {https://doi.org/10.1080/00150190701358159} {\bibfield
		{journal} {\bibinfo  {journal} {Ferroelectrics}\ }\textbf {\bibinfo {volume}
			{352}},\ \bibinfo {pages} {78--85} (\bibinfo {year} {2007})}\BibitemShut
	{NoStop}%
	\bibitem [{\citenamefont {K{\"{a}}mpfe}\ \emph {et~al.}(2014)\citenamefont
		{K{\"{a}}mpfe}, \citenamefont {Reichenbach}, \citenamefont {Schr{\"{o}}der},
		\citenamefont {Hau{\ss}mann}, \citenamefont {Eng}, \citenamefont {Woike},\
		and\ \citenamefont {Soergel}}]{Kampfe2014}%
	\BibitemOpen
	\bibfield  {author} {\bibinfo {author} {\bibfnamefont {T.}~\bibnamefont
			{K{\"{a}}mpfe}}, \bibinfo {author} {\bibfnamefont {P.}~\bibnamefont
			{Reichenbach}}, \bibinfo {author} {\bibfnamefont {M.}~\bibnamefont
			{Schr{\"{o}}der}}, \bibinfo {author} {\bibfnamefont {A.}~\bibnamefont
			{Hau{\ss}mann}}, \bibinfo {author} {\bibfnamefont {L.~M.}\ \bibnamefont
			{Eng}}, \bibinfo {author} {\bibfnamefont {T.}~\bibnamefont {Woike}},\ and\
		\bibinfo {author} {\bibfnamefont {E.}~\bibnamefont {Soergel}},\ }\bibfield
	{title} {\enquote {\bibinfo {title} {{Optical three-dimensional profiling of
					charged domain walls in ferroelectrics by Cherenkov second-harmonic
					generation}},}\ }\href {https://doi.org/10.1103/PhysRevB.89.035314}
	{\bibfield  {journal} {\bibinfo  {journal} {Physical Review B}\ }\textbf
		{\bibinfo {volume} {89}},\ \bibinfo {pages} {035314} (\bibinfo {year}
		{2014})}\BibitemShut {NoStop}%
	\bibitem [{\citenamefont {K{\"{a}}mpfe}\ \emph {et~al.}(2015)\citenamefont
		{K{\"{a}}mpfe}, \citenamefont {Reichenbach}, \citenamefont {Hau{\ss}mann},
		\citenamefont {Woike}, \citenamefont {Soergel},\ and\ \citenamefont
		{Eng}}]{Kampfe2015}%
	\BibitemOpen
	\bibfield  {author} {\bibinfo {author} {\bibfnamefont {T.}~\bibnamefont
			{K{\"{a}}mpfe}}, \bibinfo {author} {\bibfnamefont {P.}~\bibnamefont
			{Reichenbach}}, \bibinfo {author} {\bibfnamefont {A.}~\bibnamefont
			{Hau{\ss}mann}}, \bibinfo {author} {\bibfnamefont {T.}~\bibnamefont {Woike}},
		\bibinfo {author} {\bibfnamefont {E.}~\bibnamefont {Soergel}},\ and\ \bibinfo
		{author} {\bibfnamefont {L.~M.}\ \bibnamefont {Eng}},\ }\bibfield  {title}
	{\enquote {\bibinfo {title} {{Real-time three-dimensional profiling of
					ferroelectric domain walls}},}\ }\href {https://doi.org/10.1063/1.4933171}
	{\bibfield  {journal} {\bibinfo  {journal} {Applied Physics Letters}\
		}\textbf {\bibinfo {volume} {107}},\ \bibinfo {pages} {152905} (\bibinfo
		{year} {2015})}\BibitemShut {NoStop}%
	\bibitem [{\citenamefont {Kurimura}\ and\ \citenamefont
		{Uesu}(1997)}]{Kurimura1997}%
	\BibitemOpen
	\bibfield  {author} {\bibinfo {author} {\bibfnamefont {S.}~\bibnamefont
			{Kurimura}}\ and\ \bibinfo {author} {\bibfnamefont {Y.}~\bibnamefont
			{Uesu}},\ }\bibfield  {title} {\enquote {\bibinfo {title} {{Application of
					the second harmonic generation microscope to nondestructive observation of
					periodically poled ferroelectric domains in quasi-phase-matched wavelength
					converters}},}\ }\href {https://doi.org/10.1063/1.364121} {\bibfield
		{journal} {\bibinfo  {journal} {Journal of Applied Physics}\ }\textbf
		{\bibinfo {volume} {81}},\ \bibinfo {pages} {369--375} (\bibinfo {year}
		{1997})}\BibitemShut {NoStop}%
	\bibitem [{\citenamefont {Fl{\"{o}}rsheimer}\ \emph {et~al.}(1998)\citenamefont
		{Fl{\"{o}}rsheimer}, \citenamefont {Paschotta}, \citenamefont {Kubitscheck},
		\citenamefont {Brillert}, \citenamefont {Hofmann}, \citenamefont {Heuer},
		\citenamefont {Schreiber}, \citenamefont {Verbeek}, \citenamefont {Sohler},\
		and\ \citenamefont {Fuchs}}]{Florsheimer1998}%
	\BibitemOpen
	\bibfield  {author} {\bibinfo {author} {\bibfnamefont {M.}~\bibnamefont
			{Fl{\"{o}}rsheimer}}, \bibinfo {author} {\bibfnamefont {R.}~\bibnamefont
			{Paschotta}}, \bibinfo {author} {\bibfnamefont {U.}~\bibnamefont
			{Kubitscheck}}, \bibinfo {author} {\bibfnamefont {C.}~\bibnamefont
			{Brillert}}, \bibinfo {author} {\bibfnamefont {D.}~\bibnamefont {Hofmann}},
		\bibinfo {author} {\bibfnamefont {L.}~\bibnamefont {Heuer}}, \bibinfo
		{author} {\bibfnamefont {G.}~\bibnamefont {Schreiber}}, \bibinfo {author}
		{\bibfnamefont {C.}~\bibnamefont {Verbeek}}, \bibinfo {author} {\bibfnamefont
			{W.}~\bibnamefont {Sohler}},\ and\ \bibinfo {author} {\bibfnamefont
			{H.}~\bibnamefont {Fuchs}},\ }\bibfield  {title} {\enquote {\bibinfo {title}
			{{Second-harmonic imaging of ferroelectric domains in LiNbO 3 with micron
					resolution in lateral and axial directions}},}\ }\href
	{https://doi.org/10.1007/s003400050552} {\bibfield  {journal} {\bibinfo
			{journal} {Applied Physics B: Lasers and Optics}\ }\textbf {\bibinfo {volume}
			{67}},\ \bibinfo {pages} {593--599} (\bibinfo {year} {1998})}\BibitemShut
	{NoStop}%
	\bibitem [{\citenamefont {Boyd}(2003)}]{Boyd2003}%
	\BibitemOpen
	\bibfield  {author} {\bibinfo {author} {\bibfnamefont {R.~W.}\ \bibnamefont
			{Boyd}},\ }\href@noop {} {\emph {\bibinfo {title} {{Nonlinear Optics}}}}\
	(\bibinfo  {publisher} {Academic Press},\ \bibinfo {address} {London, United
		Kingdom},\ \bibinfo {year} {2003})\BibitemShut {NoStop}%
	\bibitem [{\citenamefont {R{\"{u}}sing}\ \emph {et~al.}(2019)\citenamefont {R{\"{u}}sing},
		\citenamefont {Weigel}, \citenamefont {Zhao},\ and\ \citenamefont
		{Mookherjea}}]{Rusing2019b}%
	\BibitemOpen
	\bibfield  {author} {\bibinfo {author} {\bibfnamefont {M.}~\bibnamefont
			{R{\"{u}}sing}}, \bibinfo {author} {\bibfnamefont {P.~O.}\ \bibnamefont {Weigel}},
		\bibinfo {author} {\bibfnamefont {J.}~\bibnamefont {Zhao}},\ and\ \bibinfo
		{author} {\bibfnamefont {S.}~\bibnamefont {Mookherjea}},\ }\bibfield  {title}
	{\enquote {\bibinfo {title} {{Toward 3D Integrated Photonics Including
					Lithium Niobate Thin Films: A Bridge Between Electronics, Radio Frequency,
					and Optical Technology}},}\ }\href
	{https://doi.org/10.1109/MNANO.2019.2916115} {\bibfield  {journal} {\bibinfo
			{journal} {IEEE Nanotechnology Magazine}\ }\textbf {\bibinfo {volume} {13}},\
		\bibinfo {pages} {18--33} (\bibinfo {year} {2019})}\BibitemShut {NoStop}%
	\bibitem [{\citenamefont {Hu}, \citenamefont {Ricken},\ and\ \citenamefont
		{Sohler}(2009)}]{Hu2009}%
	\BibitemOpen
	\bibfield  {author} {\bibinfo {author} {\bibfnamefont {H.}~\bibnamefont
			{Hu}}, \bibinfo {author} {\bibfnamefont {R.}~\bibnamefont {Ricken}},\ and\
		\bibinfo {author} {\bibfnamefont {W.}~\bibnamefont {Sohler}},\ }\bibfield
	{title} {\enquote {\bibinfo {title} {{Lithium niobate photonic wires}},}\
	}\href {https://doi.org/10.1364/OE.17.024261} {\bibfield  {journal} {\bibinfo
			{journal} {Optics Express}\ }\textbf {\bibinfo {volume} {17}},\ \bibinfo
		{pages} {24261} (\bibinfo {year} {2009})}\BibitemShut {NoStop}%
	\bibitem [{\citenamefont {Weigel}\ \emph {et~al.}(2018)\citenamefont {Weigel},
		\citenamefont {Zhao}, \citenamefont {Fang}, \citenamefont {Al-Rubaye},
		\citenamefont {Trotter}, \citenamefont {Hood}, \citenamefont {Mudrick},
		\citenamefont {Dallo}, \citenamefont {Pomerene}, \citenamefont {Starbuck},
		\citenamefont {DeRose}, \citenamefont {Lentine}, \citenamefont {Rebeiz},\
		and\ \citenamefont {Mookherjea}}]{Weigel2018d}%
	\BibitemOpen
	\bibfield  {author} {\bibinfo {author} {\bibfnamefont {P.~O.}\ \bibnamefont
			{Weigel}}, \bibinfo {author} {\bibfnamefont {J.}~\bibnamefont {Zhao}},
		\bibinfo {author} {\bibfnamefont {K.}~\bibnamefont {Fang}}, \bibinfo {author}
		{\bibfnamefont {H.}~\bibnamefont {Al-Rubaye}}, \bibinfo {author}
		{\bibfnamefont {D.}~\bibnamefont {Trotter}}, \bibinfo {author} {\bibfnamefont
			{D.}~\bibnamefont {Hood}}, \bibinfo {author} {\bibfnamefont {J.}~\bibnamefont
			{Mudrick}}, \bibinfo {author} {\bibfnamefont {C.}~\bibnamefont {Dallo}},
		\bibinfo {author} {\bibfnamefont {A.~T.}\ \bibnamefont {Pomerene}}, \bibinfo
		{author} {\bibfnamefont {A.~L.}\ \bibnamefont {Starbuck}}, \bibinfo {author}
		{\bibfnamefont {C.~T.}\ \bibnamefont {DeRose}}, \bibinfo {author}
		{\bibfnamefont {A.~L.}\ \bibnamefont {Lentine}}, \bibinfo {author}
		{\bibfnamefont {G.}~\bibnamefont {Rebeiz}},\ and\ \bibinfo {author}
		{\bibfnamefont {S.}~\bibnamefont {Mookherjea}},\ }\bibfield  {title}
	{\enquote {\bibinfo {title} {{Bonded thin film lithium niobate modulator on a
					silicon photonics platform exceeding 100 GHz 3-dB electrical modulation
					bandwidth}},}\ }\href {https://doi.org/10.1364/OE.26.023728} {\bibfield
		{journal} {\bibinfo  {journal} {Optics Express}\ }\textbf {\bibinfo {volume}
			{26}},\ \bibinfo {pages} {23728} (\bibinfo {year} {2018})}\BibitemShut
	{NoStop}%
	\bibitem [{\citenamefont {Bartasyte}\ \emph {et~al.}(2017)\citenamefont
		{Bartasyte}, \citenamefont {Margueron}, \citenamefont {Baron}, \citenamefont
		{Oliveri},\ and\ \citenamefont {Boulet}}]{Bartasyte2017}%
	\BibitemOpen
	\bibfield  {author} {\bibinfo {author} {\bibfnamefont {A.}~\bibnamefont
			{Bartasyte}}, \bibinfo {author} {\bibfnamefont {S.}~\bibnamefont
			{Margueron}}, \bibinfo {author} {\bibfnamefont {T.}~\bibnamefont {Baron}},
		\bibinfo {author} {\bibfnamefont {S.}~\bibnamefont {Oliveri}},\ and\ \bibinfo
		{author} {\bibfnamefont {P.}~\bibnamefont {Boulet}},\ }\bibfield  {title}
	{\enquote {\bibinfo {title} {{Toward High-Quality Epitaxial LiNbO3 and LiTaO3
					Thin Films for Acoustic and Optical Applications}},}\ }\href
	{https://doi.org/10.1002/admi.201600998} {\bibfield  {journal} {\bibinfo
			{journal} {Advanced Materials Interfaces}\ }\textbf {\bibinfo {volume} {4}},\
		\bibinfo {pages} {1600998} (\bibinfo {year} {2017})}\BibitemShut {NoStop}%
	\bibitem [{\citenamefont {Rao}\ and\ \citenamefont {Fathpour}(2018)}]{Rao2018}%
	\BibitemOpen
	\bibfield  {author} {\bibinfo {author} {\bibfnamefont {A.}~\bibnamefont
			{Rao}}\ and\ \bibinfo {author} {\bibfnamefont {S.}~\bibnamefont {Fathpour}},\
	}\bibfield  {title} {\enquote {\bibinfo {title} {{Compact Lithium Niobate
					Electrooptic Modulators}},}\ }\href
	{https://doi.org/10.1109/JSTQE.2017.2779869} {\bibfield  {journal} {\bibinfo
			{journal} {IEEE Journal of Selected Topics in Quantum Electronics}\ }\textbf
		{\bibinfo {volume} {24}},\ \bibinfo {pages} {1--14} (\bibinfo {year}
		{2018})}\BibitemShut {NoStop}%
	\bibitem [{\citenamefont {Sohler}\ \emph {et~al.}(2008)\citenamefont {Sohler},
		\citenamefont {Hu}, \citenamefont {Ricken}, \citenamefont {Quiring},
		\citenamefont {Vannahme}, \citenamefont {Herrmann}, \citenamefont
		{B{\"{u}}chter}, \citenamefont {Reza}, \citenamefont {Grundk{\"{o}}tter},
		\citenamefont {Orlov}, \citenamefont {Suche}, \citenamefont {Nouroozi},\ and\
		\citenamefont {Min}}]{Sohler2008}%
	\BibitemOpen
	\bibfield  {author} {\bibinfo {author} {\bibfnamefont {W.}~\bibnamefont
			{Sohler}}, \bibinfo {author} {\bibfnamefont {H.}~\bibnamefont {Hu}}, \bibinfo
		{author} {\bibfnamefont {R.}~\bibnamefont {Ricken}}, \bibinfo {author}
		{\bibfnamefont {V.}~\bibnamefont {Quiring}}, \bibinfo {author} {\bibfnamefont
			{C.}~\bibnamefont {Vannahme}}, \bibinfo {author} {\bibfnamefont
			{H.}~\bibnamefont {Herrmann}}, \bibinfo {author} {\bibfnamefont
			{D.}~\bibnamefont {B{\"{u}}chter}}, \bibinfo {author} {\bibfnamefont
			{S.}~\bibnamefont {Reza}}, \bibinfo {author} {\bibfnamefont {W.}~\bibnamefont
			{Grundk{\"{o}}tter}}, \bibinfo {author} {\bibfnamefont {S.}~\bibnamefont
			{Orlov}}, \bibinfo {author} {\bibfnamefont {H.}~\bibnamefont {Suche}},
		\bibinfo {author} {\bibfnamefont {R.}~\bibnamefont {Nouroozi}},\ and\
		\bibinfo {author} {\bibfnamefont {Y.}~\bibnamefont {Min}},\ }\bibfield
	{title} {\enquote {\bibinfo {title} {{Integrated Optical Devices in Lithium
					Niobate}},}\ }\href {https://doi.org/10.1364/OPN.19.1.000024} {\bibfield
		{journal} {\bibinfo  {journal} {Optics and Photonics News}\ }\textbf
		{\bibinfo {volume} {19}},\ \bibinfo {pages} {24} (\bibinfo {year}
		{2008})}\BibitemShut {NoStop}%
	\bibitem [{\citenamefont {Zhao}\ \emph {et~al.}(2020)\citenamefont {Zhao},
		\citenamefont {Ma}, \citenamefont {R{\"{u}}sing},\ and\ \citenamefont
		{Mookherjea}}]{Zhao2020}%
	\BibitemOpen
	\bibfield  {author} {\bibinfo {author} {\bibfnamefont {J.}~\bibnamefont
			{Zhao}}, \bibinfo {author} {\bibfnamefont {C.}~\bibnamefont {Ma}}, \bibinfo
		{author} {\bibfnamefont {M.}~\bibnamefont {R{\"{u}}sing}},\ and\ \bibinfo
		{author} {\bibfnamefont {S.}~\bibnamefont {Mookherjea}},\ }\bibfield  {title}
	{\enquote {\bibinfo {title} {{High Quality Entangled Photon Pair Generation
					in Periodically Poled Thin-Film Lithium Niobate Waveguides}},}\ }\href
	{https://doi.org/10.1103/PhysRevLett.124.163603} {\bibfield  {journal}
		{\bibinfo  {journal} {Physical Review Letters}\ }\textbf {\bibinfo {volume}
			{124}},\ \bibinfo {pages} {163603} (\bibinfo {year} {2020})}\BibitemShut
	{NoStop}%
	\bibitem [{\citenamefont {Wang}\ \emph {et~al.}(2018)\citenamefont {Wang},
		\citenamefont {Zhang}, \citenamefont {Chen}, \citenamefont {Bertrand},
		\citenamefont {Shams-Ansari}, \citenamefont {Chandrasekhar}, \citenamefont
		{Winzer},\ and\ \citenamefont {Lon{\v{c}}ar}}]{Wang2018f}%
	\BibitemOpen
	\bibfield  {author} {\bibinfo {author} {\bibfnamefont {C.}~\bibnamefont
			{Wang}}, \bibinfo {author} {\bibfnamefont {M.}~\bibnamefont {Zhang}},
		\bibinfo {author} {\bibfnamefont {X.}~\bibnamefont {Chen}}, \bibinfo {author}
		{\bibfnamefont {M.}~\bibnamefont {Bertrand}}, \bibinfo {author}
		{\bibfnamefont {A.}~\bibnamefont {Shams-Ansari}}, \bibinfo {author}
		{\bibfnamefont {S.}~\bibnamefont {Chandrasekhar}}, \bibinfo {author}
		{\bibfnamefont {P.}~\bibnamefont {Winzer}},\ and\ \bibinfo {author}
		{\bibfnamefont {M.}~\bibnamefont {Lon{\v{c}}ar}},\ }\bibfield  {title}
	{\enquote {\bibinfo {title} {{Integrated lithium niobate electro-optic
					modulators operating at CMOS-compatible voltages}},}\ }\href
	{https://doi.org/10.1038/s41586-018-0551-y} {\bibfield  {journal} {\bibinfo
			{journal} {Nature}\ }\textbf {\bibinfo {volume} {562}},\ \bibinfo {pages}
		{101--104} (\bibinfo {year} {2018})}\BibitemShut {NoStop}%
	\bibitem [{\citenamefont {Nikogosjan}(2005)}]{Nikogosjan2005}%
	\BibitemOpen
	\bibfield  {author} {\bibinfo {author} {\bibfnamefont {D.}~\bibnamefont
			{Nikogosjan}},\ }\href@noop {} {\emph {\bibinfo {title} {{Nonlinear Optical
					Crystals: A Complete Survey}}}}\ (\bibinfo  {publisher} {Springer},\ \bibinfo
	{address} {New York},\ \bibinfo {year} {2005})\BibitemShut {NoStop}%
	\bibitem [{\citenamefont {Shoji}\ \emph {et~al.}(1997)\citenamefont {Shoji},
		\citenamefont {Kondo}, \citenamefont {Kitamoto}, \citenamefont {Shirane},\
		and\ \citenamefont {Ito}}]{Shoji1997}%
	\BibitemOpen
	\bibfield  {author} {\bibinfo {author} {\bibfnamefont {I.}~\bibnamefont
			{Shoji}}, \bibinfo {author} {\bibfnamefont {T.}~\bibnamefont {Kondo}},
		\bibinfo {author} {\bibfnamefont {A.}~\bibnamefont {Kitamoto}}, \bibinfo
		{author} {\bibfnamefont {M.}~\bibnamefont {Shirane}},\ and\ \bibinfo {author}
		{\bibfnamefont {R.}~\bibnamefont {Ito}},\ }\bibfield  {title} {\enquote
		{\bibinfo {title} {{Absolute scale of second-order nonlinear-optical
					coefficients}},}\ }\href {https://doi.org/10.1364/JOSAB.14.002268} {\bibfield
		{journal} {\bibinfo  {journal} {Journal of the Optical Society of America
				B}\ }\textbf {\bibinfo {volume} {14}},\ \bibinfo {pages} {2268} (\bibinfo
		{year} {1997})}\BibitemShut {NoStop}%
	\bibitem [{\citenamefont {Riefer}\ \emph {et~al.}(2013)\citenamefont {Riefer},
		\citenamefont {Sanna}, \citenamefont {Schindlmayr},\ and\ \citenamefont
		{Schmidt}}]{Riefer2013b}%
	\BibitemOpen
	\bibfield  {author} {\bibinfo {author} {\bibfnamefont {A.}~\bibnamefont
			{Riefer}}, \bibinfo {author} {\bibfnamefont {S.}~\bibnamefont {Sanna}},
		\bibinfo {author} {\bibfnamefont {A.}~\bibnamefont {Schindlmayr}},\ and\
		\bibinfo {author} {\bibfnamefont {W.~G.}\ \bibnamefont {Schmidt}},\
	}\bibfield  {title} {\enquote {\bibinfo {title} {{Optical response of
					stoichiometric and congruent lithium niobate from first-principles
					calculations}},}\ }\href {https://doi.org/10.1103/PhysRevB.87.195208}
	{\bibfield  {journal} {\bibinfo  {journal} {Physical Review B}\ }\textbf
		{\bibinfo {volume} {87}},\ \bibinfo {pages} {195208} (\bibinfo {year}
		{2013})}\BibitemShut {NoStop}%
	\bibitem [{\citenamefont {Zelmon}, \citenamefont {Small},\ and\ \citenamefont
		{Jundt}(1997)}]{Zelmon2008}%
	\BibitemOpen
	\bibfield  {author} {\bibinfo {author} {\bibfnamefont {D.~E.}\ \bibnamefont
			{Zelmon}}, \bibinfo {author} {\bibfnamefont {D.~L.}\ \bibnamefont {Small}},\
		and\ \bibinfo {author} {\bibfnamefont {D.}~\bibnamefont {Jundt}},\ }\bibfield
	{title} {\enquote {\bibinfo {title} {{Infrared corrected Sellmeier
					coefficients for congruently grown lithium niobate and 5 mol{\%} magnesium
					oxide – doped lithium niobate}},}\ }\href
	{https://doi.org/10.1364/JOSAB.14.003319} {\bibfield  {journal} {\bibinfo
			{journal} {Journal of the Optical Society of America B}\ }\textbf {\bibinfo
			{volume} {14}},\ \bibinfo {pages} {3319--3322} (\bibinfo {year}
		{1997})}\BibitemShut {NoStop}%
	\bibitem [{\citenamefont {Kaneshiro}\ \emph {et~al.}(2008)\citenamefont
		{Kaneshiro}, \citenamefont {Kawado}, \citenamefont {Yokota}, \citenamefont
		{Uesu},\ and\ \citenamefont {Fukui}}]{Kaneshiro2008}%
	\BibitemOpen
	\bibfield  {author} {\bibinfo {author} {\bibfnamefont {J.}~\bibnamefont
			{Kaneshiro}}, \bibinfo {author} {\bibfnamefont {S.}~\bibnamefont {Kawado}},
		\bibinfo {author} {\bibfnamefont {H.}~\bibnamefont {Yokota}}, \bibinfo
		{author} {\bibfnamefont {Y.}~\bibnamefont {Uesu}},\ and\ \bibinfo {author}
		{\bibfnamefont {T.}~\bibnamefont {Fukui}},\ }\bibfield  {title} {\enquote
		{\bibinfo {title} {{Three-dimensional observations of polar domain structures
					using a confocal second-harmonic generation interference microscope}},}\
	}\href {https://doi.org/10.1063/1.2975218} {\bibfield  {journal} {\bibinfo
			{journal} {Journal of Applied Physics}\ }\textbf {\bibinfo {volume} {104}},\
		\bibinfo {pages} {054112} (\bibinfo {year} {2008})}\BibitemShut {NoStop}%
	\bibitem [{\citenamefont {Sandkuijl}\ \emph {et~al.}(2013)\citenamefont
		{Sandkuijl}, \citenamefont {Tuer}, \citenamefont {Tokarz}, \citenamefont
		{Sipe},\ and\ \citenamefont {Barzda}}]{Sandkuijl2013}%
	\BibitemOpen
	\bibfield  {author} {\bibinfo {author} {\bibfnamefont {D.}~\bibnamefont
			{Sandkuijl}}, \bibinfo {author} {\bibfnamefont {A.~E.}\ \bibnamefont {Tuer}},
		\bibinfo {author} {\bibfnamefont {D.}~\bibnamefont {Tokarz}}, \bibinfo
		{author} {\bibfnamefont {J.~E.}\ \bibnamefont {Sipe}},\ and\ \bibinfo
		{author} {\bibfnamefont {V.}~\bibnamefont {Barzda}},\ }\bibfield  {title}
	{\enquote {\bibinfo {title} {{Numerical second- and third-harmonic generation
					microscopy}},}\ }\href {https://doi.org/10.1364/JOSAB.30.000382} {\bibfield
		{journal} {\bibinfo  {journal} {Journal of the Optical Society of America B}\
		}\textbf {\bibinfo {volume} {30}},\ \bibinfo {pages} {382} (\bibinfo {year}
		{2013})}\BibitemShut {NoStop}%
	\bibitem [{\citenamefont {Sandkuijl}(2013)}]{Sandkuijl2013a}%
	\BibitemOpen
	\bibfield  {author} {\bibinfo {author} {\bibfnamefont {D.}~\bibnamefont
			{Sandkuijl}},\ }\emph {\bibinfo {title} {{New harmonic generation microscopy
				techniques based on focal volume modelling}}},\ \href@noop {} {Ph.D.
		thesis},\ \bibinfo  {school} {University of Toronto} (\bibinfo {year}
	{2013})\BibitemShut {NoStop}%
	\bibitem [{\citenamefont {Spychala}\ \emph
		{et~al.}(2020{\natexlab{a}})\citenamefont {Spychala}, \citenamefont
		{Mackwitz}, \citenamefont {R{\"{u}}sing}, \citenamefont {Widhalm},
		\citenamefont {Berth}, \citenamefont {Silberhorn},\ and\ \citenamefont
		{Zrenner}}]{Spychala2020a}%
	\BibitemOpen
	\bibfield  {author} {\bibinfo {author} {\bibfnamefont {K.~J.}\ \bibnamefont
			{Spychala}}, \bibinfo {author} {\bibfnamefont {P.}~\bibnamefont {Mackwitz}},
		\bibinfo {author} {\bibfnamefont {M.}~\bibnamefont {R{\"{u}}sing}}, \bibinfo
		{author} {\bibfnamefont {A.}~\bibnamefont {Widhalm}}, \bibinfo {author}
		{\bibfnamefont {G.}~\bibnamefont {Berth}}, \bibinfo {author} {\bibfnamefont
			{C.}~\bibnamefont {Silberhorn}},\ and\ \bibinfo {author} {\bibfnamefont
			{A.}~\bibnamefont {Zrenner}},\ }\bibfield  {title} {\enquote {\bibinfo
			{title} {{Nonlinear focal mapping of ferroelectric domain walls in LiNbO$_3$:
					Analysis of the SHG microscopy contrast mechanism}},}\ }\href
	{https://doi.org/10.1063/5.0025284} {\bibfield  {journal} {\bibinfo
			{journal} {Journal of Applied Physics}\ }\textbf {\bibinfo {volume} {128}},\
		\bibinfo {pages} {234102} (\bibinfo {year} {2020}{\natexlab{a}})}\BibitemShut
	{NoStop}%
	\bibitem [{\citenamefont {Sandkuijl}(2012)}]{Sandkuijl}%
	\BibitemOpen
	\bibfield  {author} {\bibinfo {author} {\bibfnamefont {D.}~\bibnamefont
			{Sandkuijl}},\ }\href {http://hdl.handle.net/1807/32992} {\enquote {\bibinfo
			{title} {{Computational code for second and third harmonic generation in
					layered media with high numerical aperture focusing}},}\ } (\bibinfo {year}
	{2012})\BibitemShut {NoStop}%
	\bibitem [{\citenamefont {Spychala}\ \emph
		{et~al.}(2020{\natexlab{b}})\citenamefont {Spychala}, \citenamefont
		{Mackwitz}, \citenamefont {Widhalm}, \citenamefont {Berth},\ and\
		\citenamefont {Zrenner}}]{Spychala2020}%
	\BibitemOpen
	\bibfield  {author} {\bibinfo {author} {\bibfnamefont {K.~J.}\ \bibnamefont
			{Spychala}}, \bibinfo {author} {\bibfnamefont {P.}~\bibnamefont {Mackwitz}},
		\bibinfo {author} {\bibfnamefont {A.}~\bibnamefont {Widhalm}}, \bibinfo
		{author} {\bibfnamefont {G.}~\bibnamefont {Berth}},\ and\ \bibinfo {author}
		{\bibfnamefont {A.}~\bibnamefont {Zrenner}},\ }\bibfield  {title} {\enquote
		{\bibinfo {title} {{Spatially resolved light field analysis of the
					second-harmonic signal of $\chi(2)$-materials in the tight focusing
					regime}},}\ }\href {https://doi.org/10.1063/1.5133476} {\bibfield  {journal}
		{\bibinfo  {journal} {Journal of Applied Physics}\ }\textbf {\bibinfo
			{volume} {127}},\ \bibinfo {pages} {023103} (\bibinfo {year}
		{2020}{\natexlab{b}})}\BibitemShut {NoStop}%
	\bibitem [{\citenamefont {Aspnes}\ and\ \citenamefont
		{Studna}(1983)}]{Apnes1983}%
	\BibitemOpen
	\bibfield  {author} {\bibinfo {author} {\bibfnamefont {D.~E.}\ \bibnamefont
			{Aspnes}}\ and\ \bibinfo {author} {\bibfnamefont {A.~A.}\ \bibnamefont
			{Studna}},\ }\bibfield  {title} {\enquote {\bibinfo {title} {{Dielectric
					functions and optical parameters of Si, Ge, GaP, GaAs, GaSb, InP, InAs, and
					InSb from 1.5 to 6.0 eV}},}\ }\href {https://doi.org/10.1103/PhysRevB.27.985}
	{\bibfield  {journal} {\bibinfo  {journal} {Physical Review B}\ }\textbf
		{\bibinfo {volume} {27}},\ \bibinfo {pages} {985--1009} (\bibinfo {year}
		{1983})}\BibitemShut {NoStop}%
	\bibitem [{\citenamefont {Volk}\ and\ \citenamefont
		{W{\"{o}}hlecke}(2008)}]{Volk2008}%
	\BibitemOpen
	\bibfield  {author} {\bibinfo {author} {\bibfnamefont {T.}~\bibnamefont
			{Volk}}\ and\ \bibinfo {author} {\bibfnamefont {M.}~\bibnamefont
			{W{\"{o}}hlecke}},\ }\href {https://doi.org/10.1007/978-3-540-70766-0} {\emph
		{\bibinfo {title} {{Lithium Niobate - Defects, Photorefraction and
					Ferroelectric switching}}}},\ \bibinfo {series} {Springer Series in Materials
		Science}, Vol.\ \bibinfo {volume} {115}\ (\bibinfo  {publisher} {Springer
		Berlin Heidelberg},\ \bibinfo {address} {Berlin, Heidelberg},\ \bibinfo
	{year} {2008})\BibitemShut {NoStop}%
	\bibitem [{\citenamefont {Wong}(2002)}]{Wong2002}%
	\BibitemOpen
	\bibinfo {editor} {\bibfnamefont {K.~K.}\ \bibnamefont {Wong}},\ ed.,\
	\href@noop {} {\emph {\bibinfo {title} {{Properties of Lithium Niobate}}}}\
	(\bibinfo  {publisher} {INSPEC, The Institution of Electrical Engineers},\
	\bibinfo {address} {London, United Kingdom},\ \bibinfo {year}
	{2002})\BibitemShut {NoStop}%
	\bibitem [{\citenamefont {Denev}\ \emph {et~al.}(2011)\citenamefont {Denev},
		\citenamefont {Lummen}, \citenamefont {Barnes}, \citenamefont {Kumar},\ and\
		\citenamefont {Gopalan}}]{Denev2011}%
	\BibitemOpen
	\bibfield  {author} {\bibinfo {author} {\bibfnamefont {S.~A.}\ \bibnamefont
			{Denev}}, \bibinfo {author} {\bibfnamefont {T.~T.~A.}\ \bibnamefont
			{Lummen}}, \bibinfo {author} {\bibfnamefont {E.}~\bibnamefont {Barnes}},
		\bibinfo {author} {\bibfnamefont {A.}~\bibnamefont {Kumar}},\ and\ \bibinfo
		{author} {\bibfnamefont {V.}~\bibnamefont {Gopalan}},\ }\bibfield  {title}
	{\enquote {\bibinfo {title} {{Probing Ferroelectrics Using Optical Second
					Harmonic Generation}},}\ }\href
	{https://doi.org/10.1111/j.1551-2916.2011.04740.x} {\bibfield  {journal}
		{\bibinfo  {journal} {Journal of the American Ceramic Society}\ }\textbf
		{\bibinfo {volume} {94}},\ \bibinfo {pages} {2699--2727} (\bibinfo {year}
		{2011})}\BibitemShut {NoStop}%
	\bibitem [{\citenamefont {Zhao}, \citenamefont {R{\"{u}}sing},\ and\
		\citenamefont {Mookherjea}(2019)}]{Zhao2019a}%
	\BibitemOpen
	\bibfield  {author} {\bibinfo {author} {\bibfnamefont {J.}~\bibnamefont
			{Zhao}}, \bibinfo {author} {\bibfnamefont {M.}~\bibnamefont {R{\"{u}}sing}},\
		and\ \bibinfo {author} {\bibfnamefont {S.}~\bibnamefont {Mookherjea}},\
	}\bibfield  {title} {\enquote {\bibinfo {title} {{Optical diagnostic methods
					for monitoring the poling of thin-film lithium niobate waveguides}},}\ }\href
	{https://doi.org/10.1364/OE.27.012025} {\bibfield  {journal} {\bibinfo
			{journal} {Optics Express}\ }\textbf {\bibinfo {volume} {27}},\ \bibinfo
		{pages} {12025} (\bibinfo {year} {2019})}\BibitemShut {NoStop}%
\end{thebibliography}
	
%
	
\end{document}